\documentclass[10pt,journal,compsoc]{IEEEtran}
\usepackage{graphics}      
\usepackage[T1]{fontenc}   
\usepackage{txfonts}
\usepackage{mathptmx}
\usepackage[pdflang={en-US},pdftex]{hyperref}
\usepackage{color}
\usepackage{booktabs}
\usepackage{subfigure}
\usepackage{multirow}
\usepackage{textcomp}
\usepackage{algorithm} 
\usepackage{algorithmic} 
\usepackage[shortlabels]{enumitem}
\setlist[enumerate, 1]{1\textsuperscript{o}}

\usepackage{microtype}        
\usepackage{ccicons}          

\usepackage{todonotes}

\newcommand{\argmin}{\arg\!\min}

\ifCLASSOPTIONcompsoc
  \usepackage[nocompress]{cite}
\else
  \usepackage{cite}
\fi
\ifCLASSINFOpdf
\else
\fi
\begin{document}
%
\title{Predicting Urban Water Quality with Ubiquitous Data}

\author{Ye Liu,
	Yuxuan Liang,
	Shuming Liu,
	~David~S.~Rosenblum,~\IEEEmembership{Fellow,~IEEE,}
	~and~Yu Zheng,~\IEEEmembership{Senior Member, ~IEEE,}
	\IEEEcompsocitemizethanks{
		\IEEEcompsocthanksitem Ye Liu and David S. Rosenblum are with the Department
		of Computer Science, School of Computing, National University of Singapore, Singapore.\protect\\
		E-mail: \{liuye, david\}@comp.nus.edu.sg
		\IEEEcompsocthanksitem Yuxuan Liang is with the Department
		of Computer Science, School of Computer Science and Technology, Xidian University, Xi'an, China. \protect\\
		E-mail: v-yuxlia@microsoft.com
		\IEEEcompsocthanksitem Shuming Liu is with the Division of Drinking Water Safety, School of Environment, Tsinghua University, China. \protect\\
		E-mail: shumingliu@tsinghua.edu.cn
		\IEEEcompsocthanksitem Yu Zheng is with Microsoft Research, Beijing, China. \protect\\
		E-mail: yuzheng@microsoft.com}
	
}

\markboth{}
{Shell \MakeLowercase{\textit{et al.}}: Bare Demo of IEEEtran.cls for Computer Society Journals}
%



\IEEEtitleabstractindextext{%
\begin{abstract}
  Urban water quality is of great importance to our daily lives. Prediction of urban water quality help control water pollution and protect human health. However, predicting the urban water quality is a challenging task since the water quality varies in urban spaces non-linearly and depends on multiple factors, such as meteorology, water usage patterns, and land uses. In this work, we forecast the water quality of a station over the next few hours from a data-driven perspective, using the water quality data and water hydraulic data reported by existing monitor stations and a variety of data sources we observed in the city, such as meteorology, pipe networks, structure of road networks, and point of interests (POIs). First, we identify the influential factors that affect the urban water quality via extensive experiments. Second, we present a multi-task multi-view learning method to fuse those multiple datasets from different domains into an unified learning model. We evaluate our method with real-world datasets, and the extensive experiments verify the advantages of our method over other baselines and demonstrate the effectiveness of our approach.
\end{abstract}

}

\maketitle

\IEEEdisplaynontitleabstractindextext

%
\IEEEpeerreviewmaketitle

\IEEEraisesectionheading{\section{Introduction}\label{sec:introduction}}

Urban water is a vital resource that affects various aspects of human, health and urban lives. People living in major cities are increasingly concerned about the urban water quality, calling for technology that can monitor and predict the water quality in real time throughout the city. Urban water quality, which serves as ``a powerful environmental determinant''  and ``a foundation for the prevention and control of waterborne diseases''~\cite{world2004guidelines}, refers to the physical, chemical and biological characteristics of a water body, and several chemical indexes (such as residual chlorine, turbidity and pH) can be used as effective measurements for the water quality in current urban water distribution systems~\cite{rossman_RCmodeling_1994}.

\begin{figure}[!h]
  \centering
  \subfigure[Stations on road network]{\label{StationRN}\includegraphics[width=0.275\textwidth]{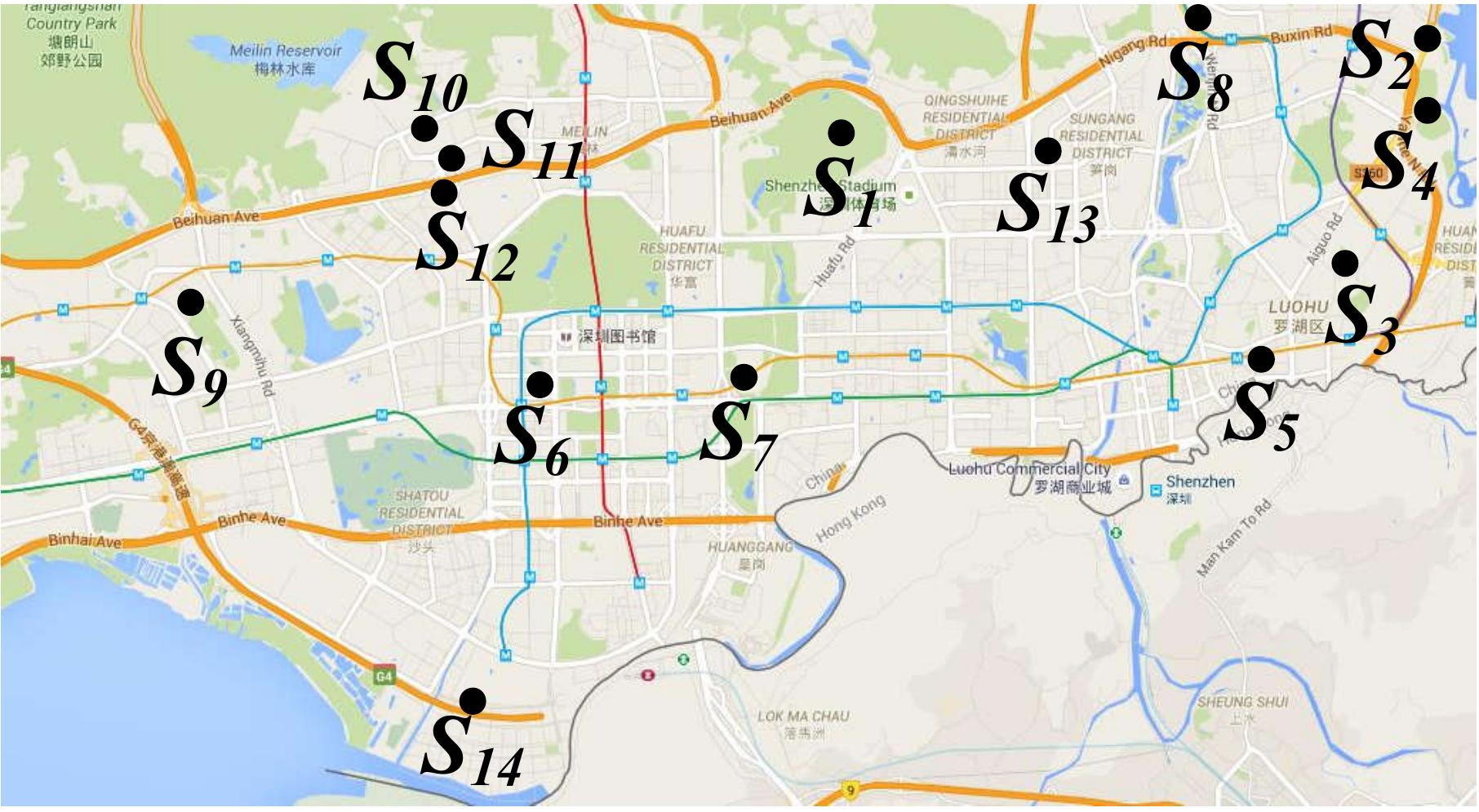}}
  \subfigure[Stations on pipe network]{\label{StationPN}\includegraphics[width=0.195\textwidth]{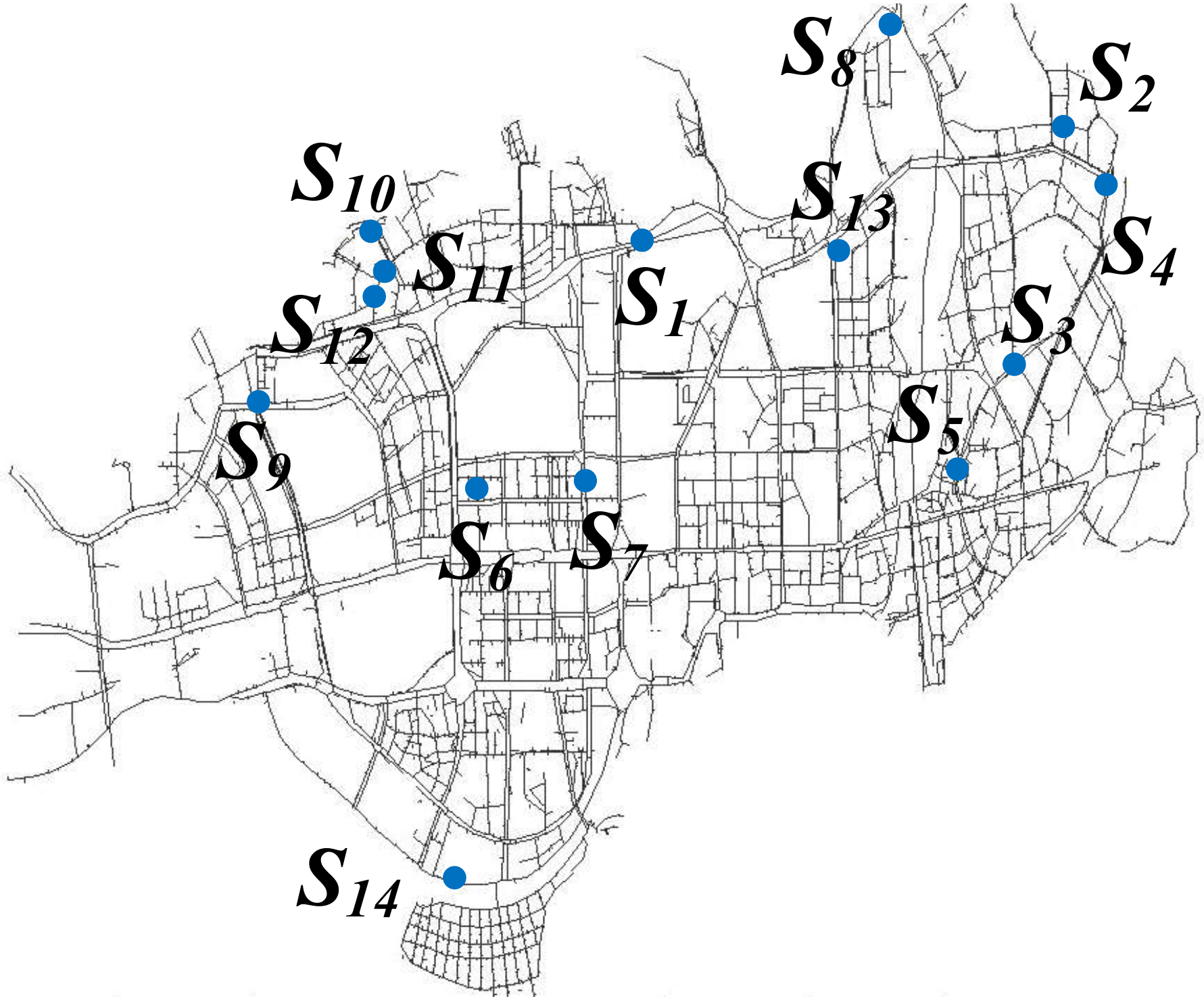}}
  \subfigure[Water quality readings in different stations]{\label{StationReading}\includegraphics[width=0.485\textwidth]{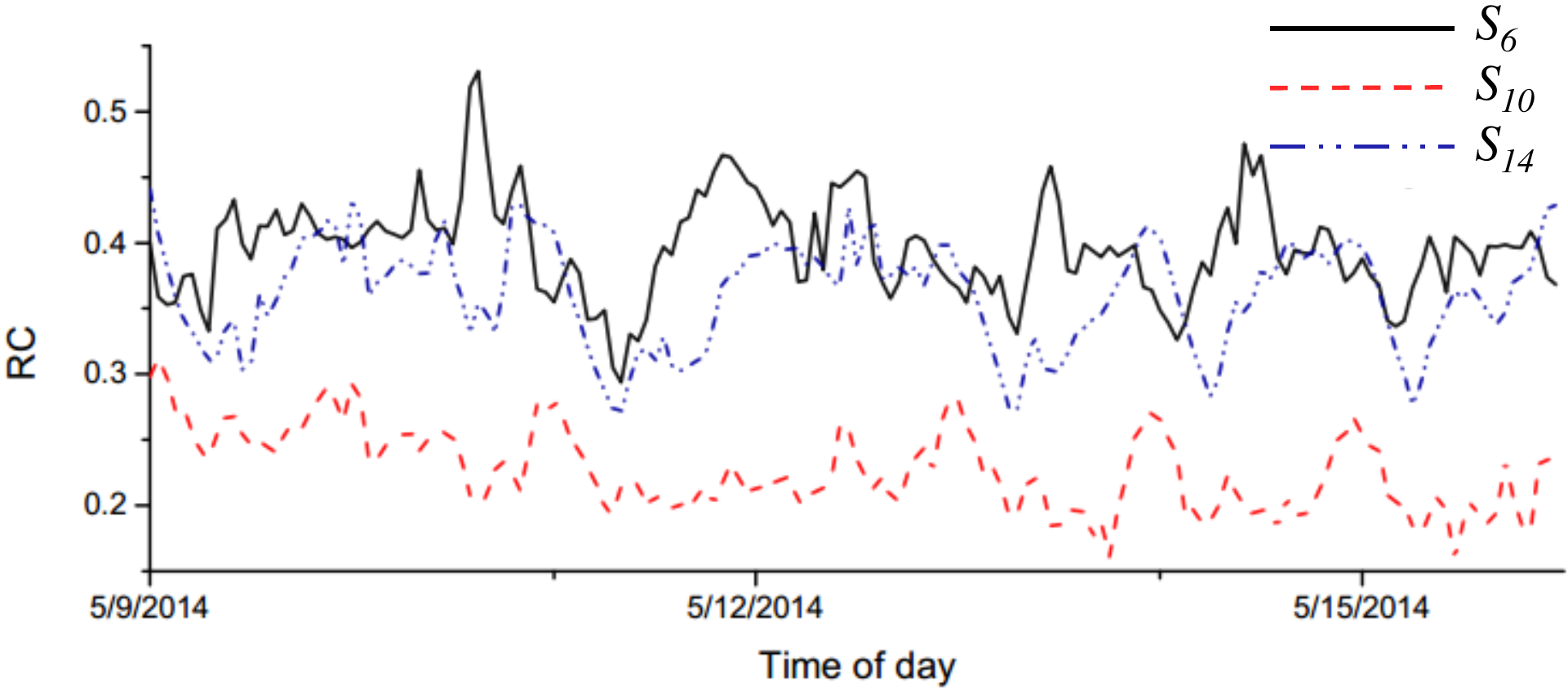}}
  \caption{Illustrations of water quality monitor stations and their readings.}\label{Introduction}
\end{figure}

With the increasing demand for water quality information, several water quality monitoring stations have been deployed throughout the city's water distribution system to provide the real-time water quality reports in a city. Figure \ref{Introduction} illustrates the water quality monitor stations that have been deployed in Shenzhen, China. Besides water quality monitoring, predicting the urban water quality plays an essential role in many urban aquatic projects, such as informing waterworks' decision making (e.g., pre-adjustment of chlorine from the waterworks), affecting governments' policy making (e.g., issuing pollution alerts or performing a pollution control), and providing maintenance suggestions (e.g., suggestions for replacements of certain pipelines).

Predicting urban water quality, however, is very challenging due to the following reasons. First, urban water quality varies by locations non-linearly and depends on multiple factors, such as meteorology, water usage patterns, land use, and urban structures. As depicted in Figure \ref{Introduction}, the water quality indexes (RC) reported by the three stations demonstrate different patterns. Exiting hydraulic model-based approaches try to model water quality from physical and chemical perspective, but such hydraulic model can hardly capture all of those complex factors. Moreover, the parameters in model are hard to get, which makes it difficult to extend to other water distribution systems. Second, as all the stations are connected through the pipeline system, the water quality among different stations are mutually correlated by several complex factors, such as attributes in pipe networks and distribution of POIs. Traditional hydraulic model-based approaches build hydraulic model for each station and ignore their spatial correlations, and thus their performance is far from satisfactory. Hence, besides identifying the influential factors, how to efficiently characterize and incorporate such relatedness poses another challenge.

Fortunately, in the era of big data~\cite{Zheng_DataFusion_TBD2015}\cite{Zheng_UrbanComputing_TIST2014}\cite{Zheng_crossDomain_GIS2015}, unprecedented data in urban areas (e.g., meteorology, POIs, and road networks) can provide complementary information to help predict the urban water quality. For example, temperature can be an indicator of water quality, with higher temperature indicating better water quality. The possible reason is that the water consumption tends to grow when temperature is high since most people may choose to take a shower, and the increased water consumption is one major cause that prevents the water quality's deterioration in the distribution systems.

To benefit from the unprecedented data in urban areas, in this paper, we predict the water quality of a station through a data-driven perspective using a variety of data sets, including water quality data, hydraulic data, meteorology data, pipe networks data, road networks data, and POIs. First, we perform extensive experiments and data analytics between the water quality and multiple potential factors, and identify the most influential ones that have an effect on the urban water quality. Second, we present a novel spatio-temporal multi-task multi-view learning (stMTMV) framework to fuse the heterogeneous data from multiple domains and jointly capture each station's local information as well as their global information into an unified learning model~\cite{MVMTL_IJCAI2016_UrbanWater}.

We summarize the contributions as follows:
\begin{itemize}
  \item \emph{Data-driven Perspective}: We present a novel data-driven approach to co-predict the future water quality among different stations with data from multiple domains. Additionally, the approach is not restricted to urban water quality prediction, but also can be applied to other multi-locations based co-prediction problem in many other urban applications.
  \item \emph{Influential Factor Identification}: We identify spatially-related (such as POIs, pipe networks, and road networks) and temporally-related features (e.g., time of day, meteorology and water hydraulics), contributing to not only our application but also the general problem of water quality prediction.
  \item \emph{Unified Learning Model}: We present a novel spatio-temporal multi-view multi-task learning framework (stMTMV) to integrate multiple sources of spatio-temporal urban data, which provides a general framework of combining heterogeneous spatio-temporal properties for prediction, and can also be applied to other spatio-temporal based applications.
  \item \emph{Real evaluation}: We evaluate our method by extensive experiments that use real-world datasets in Shenzhen, China. The results demonstrate the advantages of our method beyond other baselines, such as ARMA, Kalman filter, and ANN, and reveal interesting discoveries that can bring social good to urban life.
\end{itemize}

The rest of this paper is organized as follows: Section~\ref{Overview} overviews the framework of our method. Section~\ref{TemporalFeature} and \ref{SpatialFeature} analyze the correlations between multi-sources of urban data and the water quality. Section~\ref{Model} introduces the multi-task multi-view learning method for urban water quality prediction, and Section~\ref{Experiments} presents evaluations and visualizations. Section~\ref{RelatedWork} summarizes the related work, followed by the conclusion in the last section.

As an extension of our previous work~\cite{MVMTL_IJCAI2016_UrbanWater},  this journal version claims following contributions:
First, we focused on the data-driven perspective. Specifically, we included the insight of our methodology as well as the correlation analysis between different data with the urban water quality. The detailed correlation analysis is shown in Section \ref{TemporalFeature} and \ref{SpatialFeature}. Second, we refined the task relationship computation in our stMTMV model by figuring out the best configuration over various pipe attributes, which is achieved through the data correlation analysis in Section~\ref{GeoCorrelation}. Third, we conducted more comprehensive experiments to validate our system. For instance, we added another two popular algorithms (Kalman, ANN) as the time series prediction baselines in Section~\ref{ModelComparisonSection}. In addition, we compared the performance of our approach with other baselines over each individual station in Section~\ref{StationComparison}.

\section{Overview}\label{Overview}
\subsection{Preliminary}
\emph{\textbf{Definition 1}} (\emph{Water Quality}): Urban water quality refers to the physical, chemical and biological characteristics of a water body~\cite{rossman_RCmodeling_1994}. In current urban water distribution systems, several monitoring stations are deployed in the distribution systems to report three important quality indexes, i.e., \emph{residual chlorine}, \emph{turbidity} and \emph{pH}, in real time. The three indexes can be used as effective measurements for the water quality in current urban water distribution systems~\cite{world2004guidelines}. In this paper, we consider \emph{Residual Chlorine (RC)} as the water quality index since it ``inactivates the bacteria and some viruses that cause diarrheal disease'', and ``can protect the water from recontamination during storage''~\cite{ResidualChlorine_link_2016}, which is widely employed as the major water quality index in the environmental science~\cite{world2004guidelines}\cite{rossman_RCmodeling_1994}.

\emph{\textbf{Definition 2}} (\emph{Water Hydraulics}): Water hydraulics describe the hydraulic characteristics of the water in urban water distribution systems. It consists of two major indexes: \emph{flow} and \emph{pressure}, and can be obtained through the deployed flow and pressure sensors.

\emph{\textbf{Definition 3}} (\emph{Pipe Network}): Urban water is distributed through the pipe network, which is an underground network of interconnecting mains or pipes. A pipe network $PN$ comprises of a set of pipeline segments ${p}$ that connect between each other in the format of a graph. Each pipeline segment $p$ an undirected edge having two terminal nodes, and has several attributes, including length $p.len$, diameter $p.d$, and age $p.age$.

\emph{\textbf{Definition 4}} (\emph{Road Network}): A road network RN comprises of a set of road segments ${r}$, connecting each other in the format of a graph. Each road segment $r$ is a directed edge having two terminal nodes, a series of intermediate points between the two terminals, a length of $r.len$.

\subsection{Insight}
\begin{figure}[!h]
  \centering
  \includegraphics[width=0.49\textwidth]{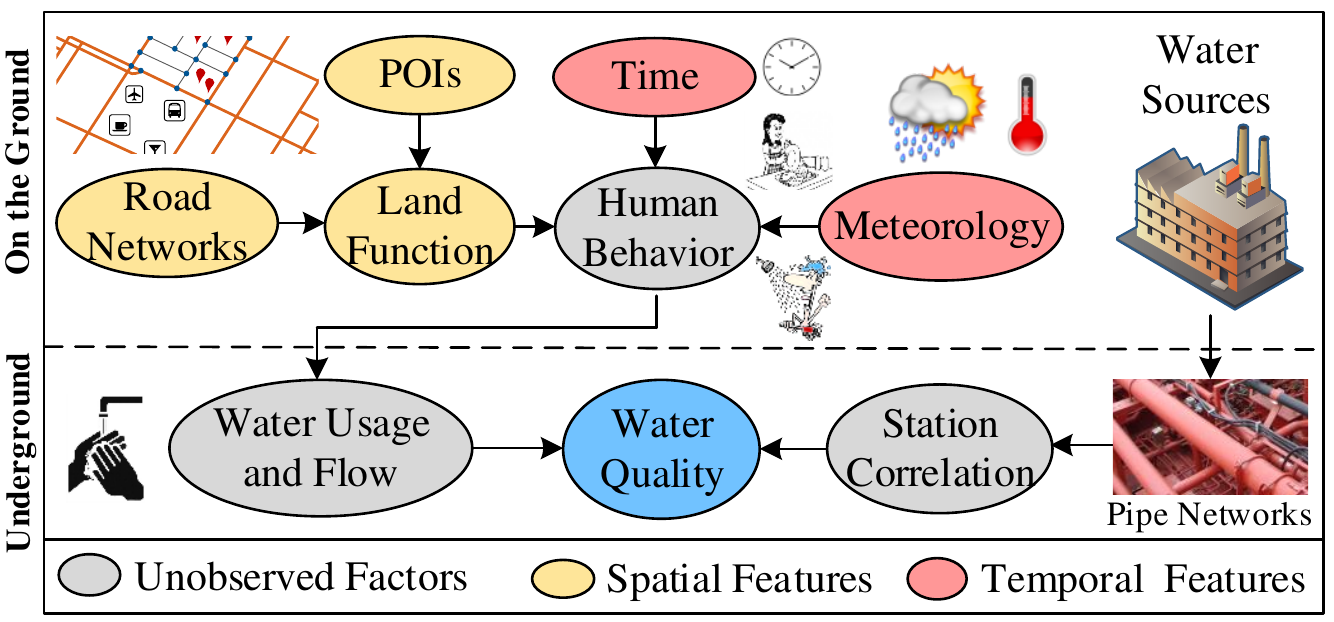}\\
  \caption{Illustration of interrelationship among influential factors for the urban water quality prediction.}\label{AnalyticsOverview}
\end{figure}

In this section, we present the motivation of our method by analyzing the correlations of urban water quality with its influential factors, including hydraulic characteristics, road network, pipe network, meteorology, and point of interest (POI). Figure \ref{AnalyticsOverview} depicts the interrelationship of various influential factors for the urban water quality, where pink nodes are temporal factors, yellow nodes denote spatial factors, and gray nodes represent unobserved factors. Generally, those factors can be categorized as direct and indirect contributing factors. Those that have intermediate impacts on the water quality are direct contributing factors. For example, hydraulic characteristics and pipe network have direct influence on the water quality since water from the service reservoirs is distributed to users via extensive networks of water mains and pipe corrosion or flow velocities can cause water quality to deteriorate in a direct way.
Hence, hydraulic characteristics and pipe network are main influencing factors for the water quality. Besides direct factors, those that do not have direct influences on the water quality are considered as indirect contributing factors. Take time as an example, time plays an important role in the prediction of water quality, even if it is not directly related to it. One reason is that time is a key factor that affect human behavior, which has an effect on the water quality by influencing the water usage patterns. Similarly, meteorology, road network, POIs are indirect contributing factors for the water quality.

\subsection{Framework}
\begin{figure}[!h]
  \centering
  \includegraphics[width=0.49\textwidth]{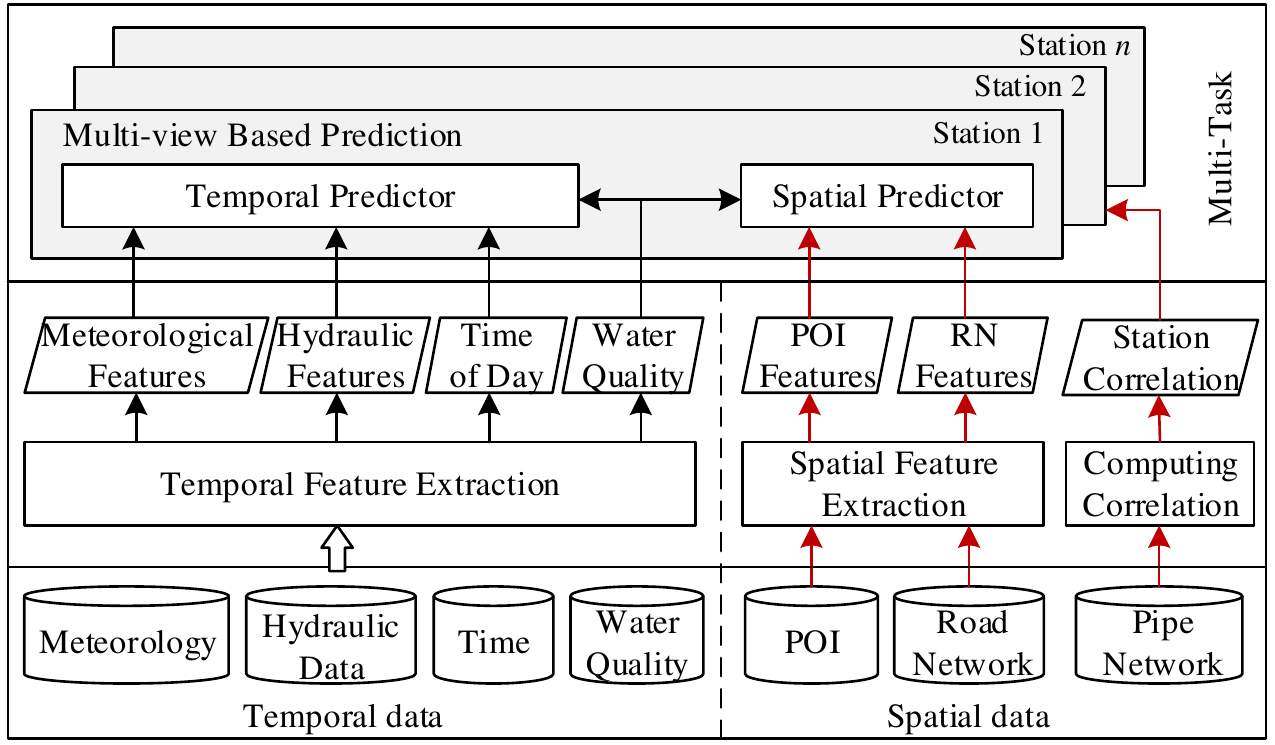}\\
  \caption{The framework of our approach.}\label{SystemOverview}
\end{figure}

Figure \ref{SystemOverview} presents the framework of our approach, which consists of three major components. 1) Feature extraction. We extract meteorological features, hydraulic features, and time of day from the temporal data in each station, and we also extract road network features and POIs features from each station's spatial data. 2) Multi-view based prediction. After extracting spatial- and temporal-related features, spatial and temporal views are constructed by for each station, and multi-view based prediction is employed to capture the local information within a station (node) and combine its spatial and temporal features. 3) Multi-task based prediction. As the water quality among stations are mutually correlated through the complex water distribution system, we thus can co-predict the water quality over all stations by capturing their spatial correlations. Therefore, multi-task based prediction is used to jointly aligns the predictions among different stations (nodes) by incorporating the their similarities in the pipe networks. We will detail the three major components in the following sections respectively.

\section{Temporal Feature Extraction}\label{TemporalFeature}

\subsection{Hydraulic Features}
It has been well studied in the area of environmental research that water hydraulics, such as water flow and pressure, are closely related to water quality, and the measurement of water hydraulics is an important component of most water quality monitoring projects~\cite{barkdoll2003effect}\cite{castro2003chlorine}\cite{mays1999water}\cite{rossman1996numerical}. As the urban water travels through the water distribution system, the flow reflects its velocity as well as the water age in the system, where the water age is one major contributing factor in water quality deterioration within the distribution system~\cite{castro2003chlorine}\cite{grayman1988modeling}. The reason is that the residual chlorine in water normally reacts with organic material and pipe walls, and thus residual disinfectant concentration will diminish as water lingers in the distribution system. If treated water stays in the system a longer time before it reaches users, the disinfectant concentration may not be strong enough after undergoing various chemical reactions.

\begin{figure}[!h]
  \centering
  \subfigure[flow-pressure-RC]{\label{Flow-Press-RC}\includegraphics[width=0.24\textwidth]{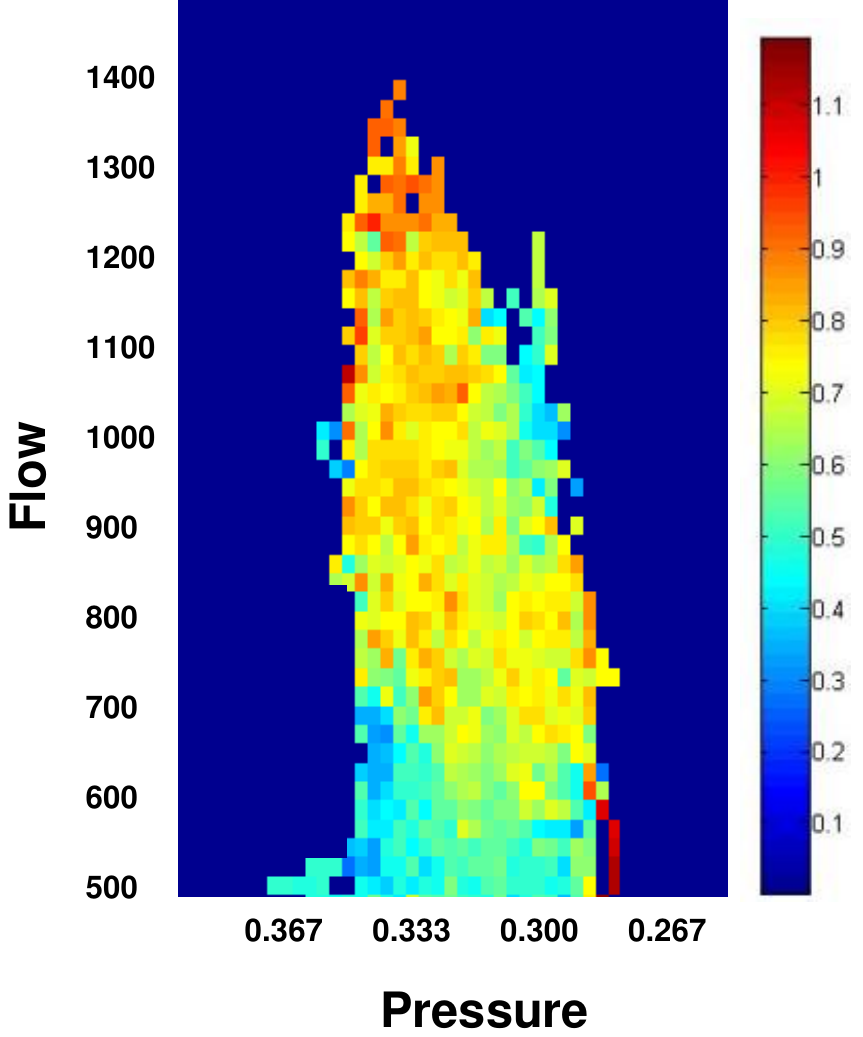}}
  \subfigure[turbidity-pH-RC]{\label{TR-PH-RC}\includegraphics[width=0.24\textwidth]{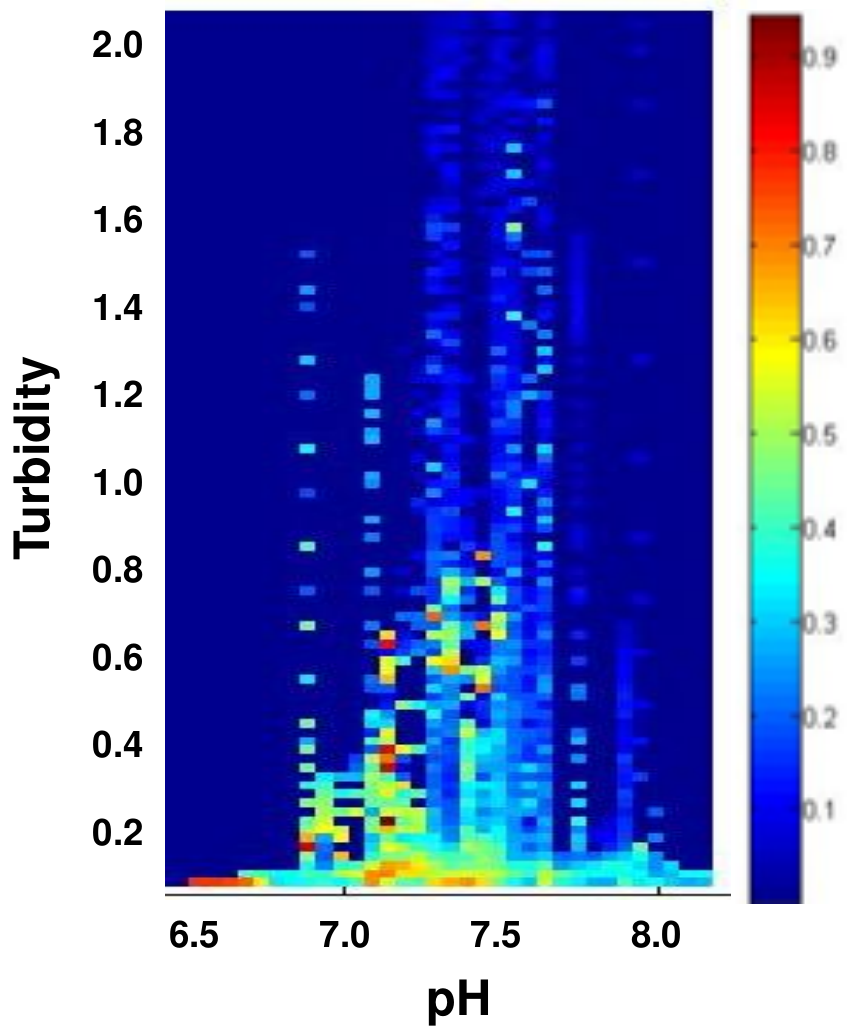}}
  \caption{Correlation between water quality and water hydraulics.}\label{PH-TR-RC}
\end{figure}

This phenomenon can also be observed from data as illustrated in Figure \ref{Flow-Press-RC}, where color, x- and y-axis denote {RC}, pressure and flow, respectively. From Figure \ref{Flow-Press-RC}, it can be seen that when water travels with a higher speed in the distribution system, i.e., $flow>700 m^{3}/h$, more instances of high density of {RC} occurred. On the contrary, the smaller the flow ($flow<700 m^{3}/h$) is, the lower density of {RC} would be. Clearly, this verifies that there exists
relatedness between flow and {RC} and the negative correlation is very intuitive to understand given the fact demonstrated above. Although the pressure does not have intermediate impact on water age, it has positive effects on the water flow, which, in turn, affects the water age as well as the water quality. An interesting observation is that a larger pressure indicates a higher density of {RC} in the water while a smaller one has a very high probability of resulting in a lower density of {RC}, as depicted in Figure \ref{Flow-Press-RC}. The results well verify the assumption that pressure is another contributing factor of water quality. Thus, water hydraulic characteristics (flow and pressure) are closely correlated with the water quality, and can also be used as the general indicators of water quality.

As flow and pressure are both time series signals, we extract several time series features to capture the characteristics of the signal comprehensively. In particular, we use the latest 12 hours water hydraulic data and extract statistical features (mean, variance, maximum, minimum, skewness and kurtosis), and time series features (autocorrelation, PAA~\cite{Lin_PAA_SIGMOD2007}, PLA~\cite{Luo_PLA_ICDE2015}) for each of the time series in the hydraulic data. Moreover, we also extract frequency related features (FFT~\cite{brigham1974fast_FFT} and DWT~\cite{burrus1997introduction_DWT}) for each of the time series above, where we only use the top $3$ coefficients and discard others.

\subsection{Meteorological Features and Time of Day}
\begin{figure}[!h]
  \centering
  \includegraphics[width=0.49\textwidth]{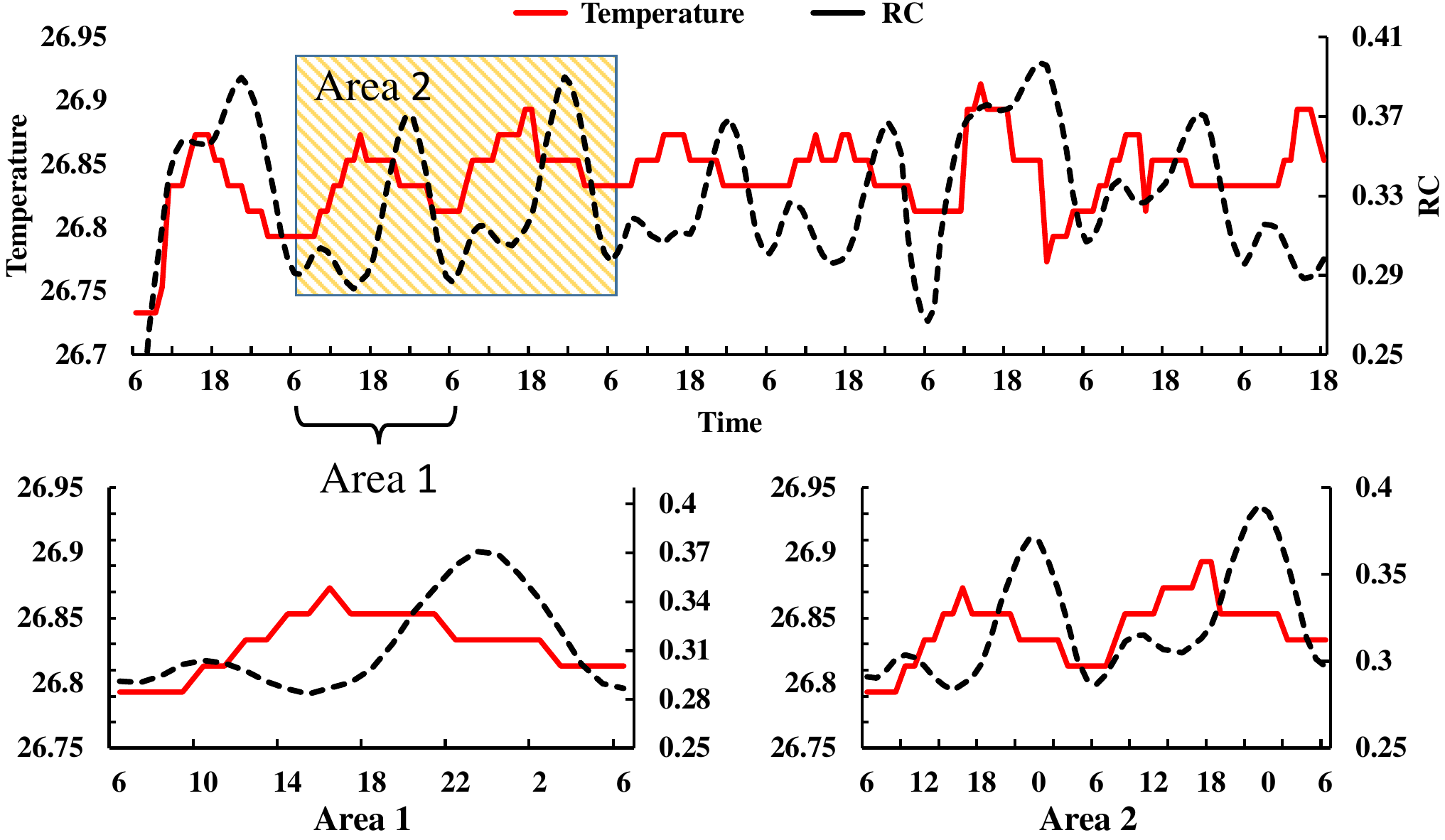}\\
  \caption{Correlation between temperature and RC.}\label{Temperature-RC}
\end{figure}

The concentration of RC in urban water system is influenced by meteorology indirectly as shown in Figure \ref{AnalyticsOverview}, where meteorology denotes temperature, humidity, barometer pressure, wind speed, and weather. In particular, the meteorology has direct impacts on human behavior, and it thus affects the water usage patterns, which influence the water quality as a consequence. Take the temperature as an example, the temperature has positive correlation with {RC} and a high temperature is always accompanied with a high concentration of {RC}. This is partially due to the fact that the high water consumption is growing when the temperature goes high, which in turn has influence on the water quality. This pattern can be observed from the case example of ``Area 2'' in Figure \ref{Temperature-RC}. In particular, the second day shows a higher temperature as compared to the first day, and the concentration of RC on the second day is slightly higher as well.

Time of day is another contributing cause for the urban water quality due to its strong correlation with human behavior. For instance, 6am to 10am could be the rush hour or morning peak in most cities, and people in cities normally get up, wash or even take a shower at that time. The increased water consumption leads to a higher concentration of {RC} in the water system. From Figure \ref{Temperature-RC}, it can be seen that the concentration of {RC} has a slightly growing trend in that period of time. Similar patterns can also be observed from 6pm to 11pm since people usually go home after 6pm and do some cooking or washing up in the evening. On the contrary, the concentration of {RC} declines gradually after 11am. It is because people go to sleep after 11pm that the water consumption is reduced as well.

In summary, meteorology and time of day can provide complementary information of human behavior, helping predict the water quality. In particular, we extract temperature, humidity, barometer pressure, wind speed, and weather as the meteorological features, and time of day as the time feature.

\subsection{Water Quality}
A variety of water quality-related studies have been performed and several research efforts have demonstrated the interrelationship among the three water quality indexes ({RC, turbidity, pH})~\cite{lechevallier1981effect}\cite{monteiro2014modeling}\cite{world2004guidelines}\cite{rossman_RCmodeling_1994}. As a measure of the cloudiness of the water, turbidity was found to be associated with the total organic carbon in the water, which has been shown to interfere with the free chlorine residual by chemical reactions~\cite{lechevallier1981effect}. In particular, turbidity normally exhibits opposite trend with {RC} since the chemical reactions of {RC} with pipe and bulk fluid will consume {RC} and increase the turbidity in water~\cite{castro2003chlorine}. This negative correlation can also be observed from data as shown in Figure \ref{TR-PH-RC}. Figure \ref{TR-PH-RC} shows the correlation between the {RC} and the the other indexes ({turbidity} and {pH}), using the data we collected from August to Dec. 2013 in Shenzhen, where color, x- and y-axis denote {RC}, {pH} and {turbidity}, respectively. Apparently, a high density of {RC} will lead to a low turbidity in the water, and the high turbidity in water occurs as a result of consumption of {RC} and causes low density of {RC} in water. Moreover, it can be observed that pH around $7.0$ would result in a high density of {RC}. In short, {turbidity} and {pH} are correlated with {RC}, and can be employed as discriminative features for water quality ({RC}) predictions.

To fully capture the information from water quality data, we employ similar feature extraction procedure by using the latest 12 hours water quality data and extracting statistical features (mean, variance, maximum, minimum, skewness and kurtosis), frequency related features (FFT~\cite{brigham1974fast_FFT} and DWT~\cite{burrus1997introduction_DWT}) and time series features (autocorrelation, PAA~\cite{Lin_PAA_SIGMOD2007}, PLA~\cite{Luo_PLA_ICDE2015}) for each of the time series in the historical water quality data (RC, turbidity, and pH).

\section{Spatial Feature Extraction}\label{SpatialFeature}
\subsection{Road Network Features}
\begin{figure}[!h]
  \centering
  \includegraphics[width=0.49\textwidth]{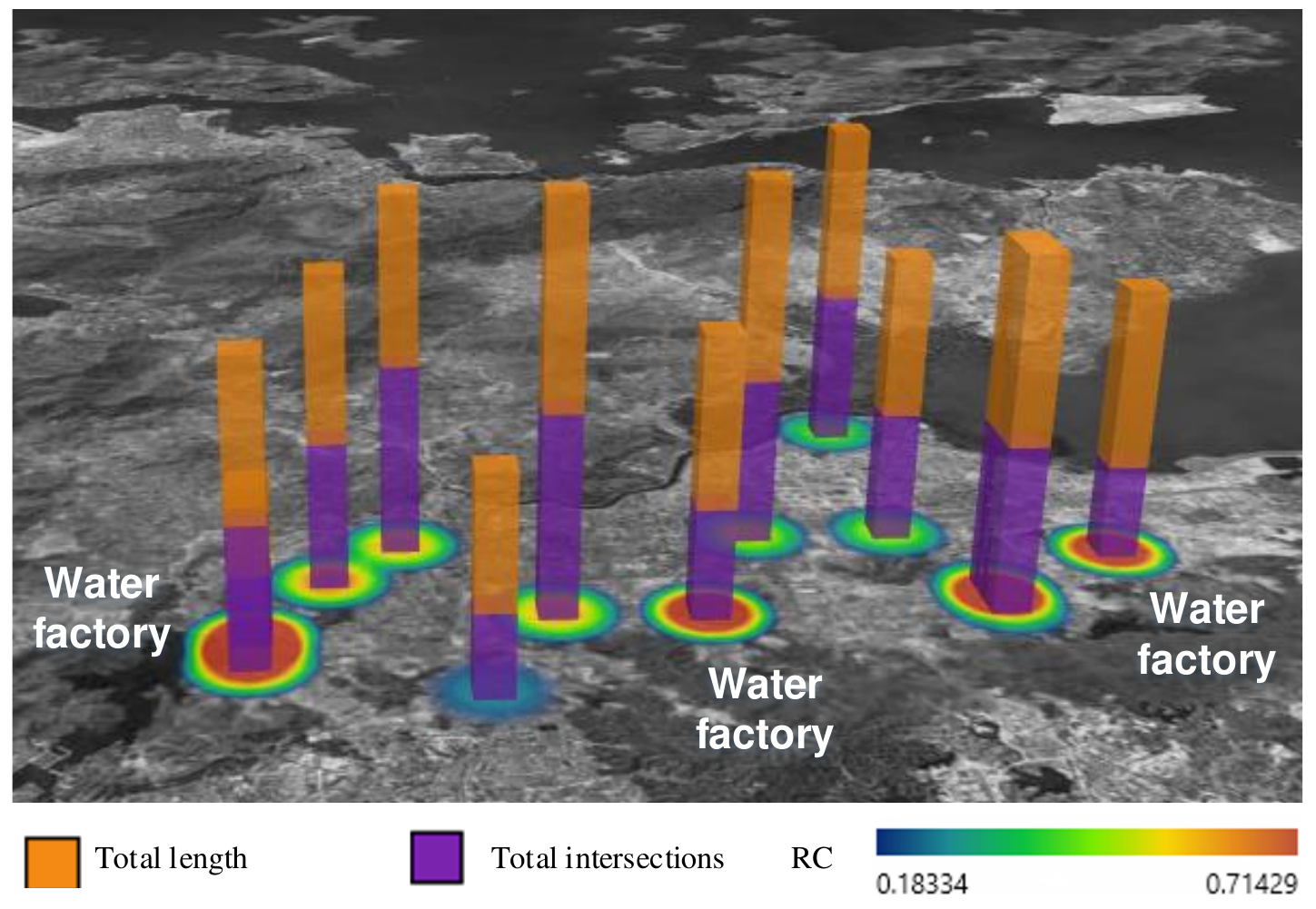}\\
  \caption{Correlation between road network features and RC.}\label{RoadNetwork-RC}
\end{figure}

As most of the pipelines are constructed underneath the roads, the structure of a road network in an area, like the number of intersections and the total length of road segments, reflects the structure of the pipe network. Thus, it also has correlation with the region's water quality, which can be used as complementary information for water quality prediction. Figure \ref{RoadNetwork-RC} shows the correlation between water quality in a station and a few road network features (e.g. the total length of road segments, and the number of intersections). Each pillar represents one water quality monitoring station, and its height denotes the feature strength. Different colors of the circle around the station stand for different concentrations of RC in the station, e.g. orange denotes a high concentration of RC in a station. So, each pillar in Figure \ref{RoadNetwork-RC} shows the concentration of RC in a station with respect to two road network features. As demonstrated in the height of the pillar and color of the circle around it, we can clearly see that the more intersections are around a station, the more orange circles occur. We also find a similar pattern with respect to the total length of roads in Figure \ref{RoadNetwork-RC}.

Thus, we identify two features for each station based on the road network structure: 1) the total length $f_{rl}$, 2) the number of intersections $f_{rs}$ in the station's affecting region.

\subsection{POIs Features}
The information on POIs in a region, such as the number of POIs in different categories and the density of POIs, indicates the function of the region as well as the water usage patterns in the region, which has an impact on the region's hydraulic conditions (e.g., flow, pressure) and greatly affect water quality as a consequence. For example, if many factories or car repair workshops appears in a region, the consumption of water in that area tends to be high. This leads to a high concentration of RC in the water since the water stays much shorter in the pipeline system and the short time stay can hardly cause chlorine residual to fade. On the contrary, the demand for water in places like parks is usually low, and thus the long detention times in the distribution system result in degraded water quality.

\begin{figure}[!h]
  \centering
  \includegraphics[width=0.49\textwidth]{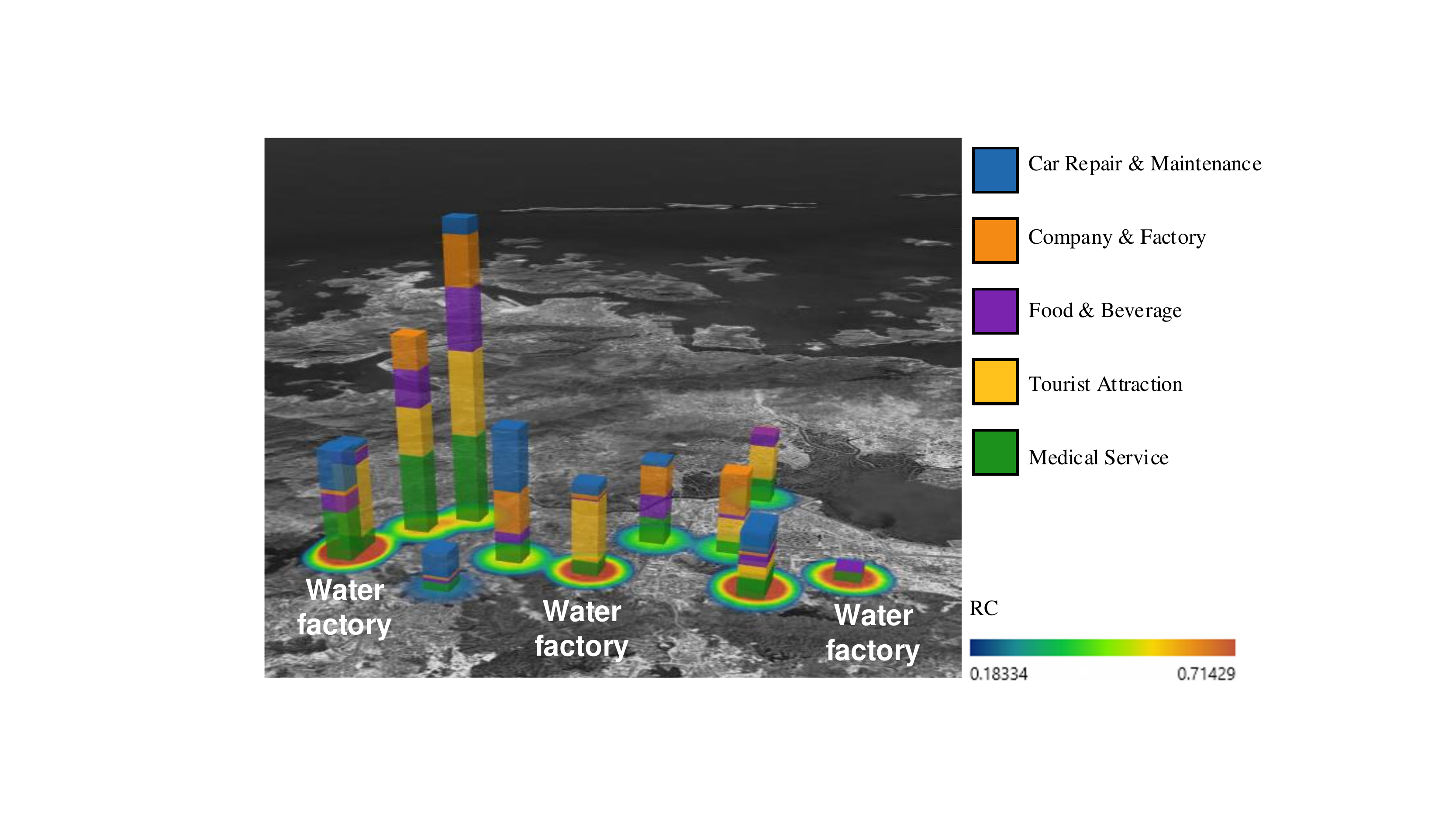}\\
  \caption{Correlation between the density of POIs and RC.}\label{POI-RC}
\end{figure}

Figure \ref{POI-RC} illustrates the correlation between the concentration of RC and the density of POIs among different monitoring stations. Each pillar denotes a water quality monitoring station, and its height represents the density of POIs around the station. The concentration of RC for each station is expressed by the color of the circle around it, where the orange one denotes a high concentration of RC and the blue one is the opposite. Among $14$ monitoring stations, three of them are waterworks. Hence, the concentration of RC in the treated water is high.

As depicted in Figure \ref{POI-RC}, the density of POIs demonstrates positive correlation with the concentration of RC. In particular, it can be observed that a high density of POIs can cause the concentration of RC to be high in that region. An exceptional case is the station that near the waterworks, which does not demonstrate this correlation strongly. However, those exceptional cases provide evidences of the geographical similarity among stations squarely. So, POI data can also be treated as complementary information to help predict the water quality of a station. As compared to the distributions in road network, there are still some differences between them, and each piece of data may only tell us a part of the panoramic view of urban water quality. Thus, that is the reason why we need to incorporate multiple data sources.

\section{Multi-task Multi-view Learning Model}\label{Model}
%

\subsection{Notation}
We first define some notations. In particular, we use bold capital letters (e.g., \textbf{X}) and bold lowercase letters (e.g., \textbf{x}) to denote matrices and vectors, respectively. We employ non-bold letters (e.g., x) to represent scalars, and Greek letters (e.g., $\lambda$) as parameters. Unless stated, otherwise, all vectors are in column form. Let us assume that we have $M$ nodes for the water quality prediction, and each node $l$ is described by its spatial view $ \textbf{X}_{l}^{s}\in \mathbb{R}^{N_{l}\times D_{s}} =[\textbf{x}_{l,1}^{s}, \textbf{x}_{l,2}^{s}, \ldots, \textbf{x}_{l,N_{l}}^{s}]^{T}$ and temporal view $ \textbf{X}_{l}^{t}\in \mathbb{R}^{N_{l}\times D_{t}} = [\textbf{x}_{l,1}^{t}, \textbf{x}_{l,2}^{t}, \ldots, \textbf{x}_{l,N_{l}}^{t}]^{T}$, where $ \textbf{x}_{l,i}^{s} \in \mathbb{R}^{D_{s}}$ and $ \textbf{x}_{l,i}^{t} \in \mathbb{R}^{D_{t}}$ denote the spatial feature and temporal feature extracted from the node $l$ at time point $i$. $N_{l}$ is the number of samples at node $l$, and $ D_{s}$ and $ D_{t}$ is the feature dimension of the spatial view and temporal view, respectively. The whole feature matrix at node $l$ can be written as $ \textbf{X}_{l} = [\textbf{X}_{l}^{s}, \textbf{X}_{l}^{t}] \in \mathbb{R}^{N_{l}\times D}$, where $ D = D_{s}$+$D_{t}$. The target vector at node $l$ is denoted as $ \textbf{y}_{l} = \{\textbf{y}_{l,1}, \textbf{y}_{l,2}, \ldots, \textbf{y}_{l,N_{l}}\}\in \mathbb{R}^{N_{l}}$, which represents the water quality of node $l$ observed at the discrete time points $ 1,2,\ldots, N_{l}$. $ N = \sum_{l=1}^{M}N_{l}$ is the total number of samples over all nodes.

\subsection{Technical Formulation}
Technically, our multi-task multi-view learning model consists of two components: multi-view learning and multi-task learning, and each component plays a different role in our model. In particular, multi-view learning aims to combine the two different views within a station, which helps boost the local predictive performance. The multi-task learning captures the pipe network correlations among different stations and aligns their predictions, and this alignment can improve the co-predictive performance over all stations. By combining the multi-view learning and multi-task learning components, we can get an unified spatio-temporal multi-task multi-view learning model (stMTMV) to perform the water quality co-prediction problem. Moreover, the dimension of features for prediction is usually very high, but not all features are sufficiently discriminative for water quality prediction. To select a common set of discriminative features among all tasks, we employ a group Lasso penalty~\cite{yuan2006model_GroupLasso}, which can identify the top sharing features automatically. The overall objective function for the spatio-temporal multi-task multi-view learning model (stMTMV) can be formulated as
\begin{eqnarray}\label{PredictionObjectiveFunction}
\nonumber
\mathop{\min}_{\textbf{W}} && \frac{1}{2}\sum_{l = 1}^{M}\|\textbf{y}_{l}-\frac{1}{2}\textbf{X}_{l}\textbf{w}_{l}\|_{2}^{2} + {\lambda}\sum_{l = 1}^{M}\|\textbf{X}_{l}^{s}\textbf{w}_{l}^{s}-\textbf{X}_{l}^{t}\textbf{w}_{l}^{t}\|_{2}^{2} \\
&& + {\gamma}\sum_{l,m = 1}^{M}C_{l,m}\|\textbf{w}_{l} - \textbf{w}_{m} \|_{2}^{2} + \theta\|\textbf{W}\|_{2,1},
\end{eqnarray}
where the first term is the loss function, the second term is the multi-view learning component, the third term is multi-task learning component and $\lambda, \gamma, \theta$ are regularization parameters. The $\ell_{2,1}$-norm of a matrix $\textbf{W}$ is defined as $\|\textbf{W}\|_{2,1}=\sum_{i=1}^{D}\sqrt{\sum_{j=1}^{M}W_{ij}^2}$. In particular,  $\ell_{2,1}$-norm applies an $\ell_{2}$-norm to each row of $\textbf{W}$ and these $\ell_{2}$-norms are combined through an $\ell_{1}$-norm. As we assume that only a small set of features are predictive for prediction task, the $\ell_{2,1}$-norm encourages all tasks to select a common set of features and thereby plays the role of group feature selection.

\subsection{Multi-view Learning}
Multi-view learning has been proposed to leverage the information from diverse domains or from various feature extractors, and combining the heterogeneous properties from different views can better characterize objects and achieve promising performance~\cite{zhang2013multi_IJCAI2013}\cite{Zheng_DataFusion_TBD2015}\cite{MVMTL_AAAI2016_Career}\cite{sun2013survey_MVL}\cite{xu2013survey_MVL}. In this work, as there are various spatial- and temporal data around a water quality station, the information in a station can be represented by two separate views, i.e., spatial view and temporal view, thus, combining the two different views within a station can achieve better performance.

\subsubsection{Spatio-temporal Views Construction}
First, we construct the spatio-temporal views by extracting spatial- and temporal-related features for each station from spatial and temporal datasets, respectively. In particular, the temporal view of each station is constructed by incorporating its temporal information, such as historical water quality data (RC, turbidity, pH), historical hydraulic data (flow, pressure), and meteorological data, and extracting temporal features from above data. The spatial view of each station is constructed by integrating its spatial inferential factors, such as pipe network, road network, and POIs, and extracting spatial features as well as the aggregated surrounding temporal features, which encodes its spatial information as well.

\subsubsection{Technical Formulation}
Next, we will get into the technical formulation of multi-view learning. The prediction at each node $l$ consists of spatial prediction and temporal prediction, i.e., $ f_{l}^{s}(\textbf{X}_{l}^{s}) = \textbf{X}_{l}^{s}\textbf{w}_{l}^{s}$ for spatial prediction and $ f_{l}^{t}(\textbf{X}_{l}^{t}) = \textbf{X}_{l}^{t}\textbf{w}_{l}^{t}$ for temporal prediction, where $ \textbf{w}_{l}^{s}\in \mathbb{R}^{D_{s}}$ and $ \textbf{w}_{l}^{t}\in \mathbb{R}^{D_{t}}$ denote the linear mapping function for the node $l$ with spatial view and temporal view, respectively. In this paper, we use the linear function for simplicity. However, our model can be extended to other convex and smooth non-linear prediction functions. Without prior knowledge on the contributions of spatial and temporal view, we assume that both contribute equally. Thus, the final prediction model of both spatial and temporal view for node $l$ is obtained by the following late fusion~\cite{zhang2012inductive_regMVMT_KDD2012}\cite{MVMTL_AAAI2016_Career}\cite{sun2013survey_MVL}:
\begin{equation}\label{prediction_task}
  f_{l}(\textbf{X}_{l}) = \frac{1}{2}(f_{l}^{s}(\textbf{X}_{l}^{s})+f_{l}^{t}(\textbf{X}_{l}^{t}))=\frac{1}{2}(\textbf{X}_{l}^{s}\textbf{w}_{l}^{s}+\textbf{X}_{l}^{t}\textbf{w}_{l}^{t})=\frac{1}{2}\textbf{X}_{l}\textbf{w}_{l},
\end{equation}
where $ \textbf{w}^{l}\in \mathbb{R}^{D}$ is the weight vector for node $l$. The weight matrix over $M$ nodes is denoted as $ \textbf{W} = [\textbf{w}_{1}, \textbf{w}_{2}, \ldots, \textbf{w}_{M}]\in \mathbb{R}^{D\times M}$.

Information distributed in spatial and temporal views in fact describes the inherent characteristics of the same node from various aspects, we thus can reinforce the learning performance of individual views by enforcing the agreement on the their prediction results. Considering the least-squares loss function, we can define the following objective function~\cite{zhang2012inductive_regMVMT_KDD2012}\cite{MVMTL_AAAI2016_Career}\cite{xu2013survey_MVL}:
\begin{equation}\label{sourceConsistence}
   \mathop{\min}_{\textbf{W}} \frac{1}{2}\sum_{l = 1}^{M}\|\textbf{y}_{l} - \frac{1}{2}\textbf{X}_{l}\textbf{w}_{l}\|_{2}^{2} + {\lambda}\sum_{l = 1}^{M}\|\textbf{X}_{l}^{s}\textbf{w}_{l}^{s} - \textbf{X}_{l}^{t}\textbf{w}_{l}^{t} \|_{2}^{2},
\end{equation}
where $\lambda$ is a regularization parameter, $\|\textbf{x}\|_{2}=\sqrt{\sum_{i}x_{i}^2}$ is the $\ell_{2}$-norm. In a real pipeline system, each node is not only affected by its local information, but also affected by the information from its neighbors or other nodes.

\subsection{Multi-task Learning}
Multi-task learning is a learning paradigm that jointly learns multiple related tasks and has demonstrated its advantages in many urban applications~\cite{YeJieping_KDD2015_MTLSpatial,Zheng_UrbanComputing_TIST2014}. In particular, it is more effective in handling those with insufficient training samples~\cite{Liu_IJCAI2015_MultiTask}\cite{ZhangYu_UAI_MTL_2010}\cite{YeJieping_KDD2015_MTLSpatial}\cite{zhang2013multi_IJCAI2013}\cite{zhou2011multi_MTL}\cite{zhou2012modeling_MTL}. In this work, the water quality prediction in each node is aligned with a task. As the urban water is distributed across the water distribution systems, and hence the water quality among stations are mutually correlated. Therefore, we can co-predict the water quality over all stations by capturing the spatial correlations among them, and these correlations are pre-computed by the structure of pipe network.

\subsubsection{Pipe Network Correlation}\label{GeoCorrelation}
The water quality among stations are mutually correlated through the complex water distribution system and thus there exist spatial correlations among them. Therefore, how to capturing those spatial correlations becomes a quite important question. As water is distributed through the water distribution system, the pipe network becomes the major contributing factor for the spatial correlations. The effects of various attributes of water pipes on the water quality have also been explored in the environmental science~\cite{frateur1999free}\cite{lechevallier1993examining}\cite{madlala2001xylanase}, and the impact of water pipes on the water quality varies from system to system, depending on the attributes of each pipe as well as the whole pipe network configuration~\cite{clark2005characterizing}\cite{frateur1999free}\cite{lechevallier1993examining}\cite{madlala2001xylanase}\cite{niquette2000impacts}. More specifically, among various attributes, the \textit{length}, \textit{diameter}, and \textit{age} of a pipe are potential water quality deterioration factors that play an important role in affecting the water age, which consequently influences water quality. In this paper, we computed the spatial correlations by taking account of various pipe attributes and the pipe network configurations. 

\begin{figure}[!h]
	\centering
	\subfigure[Pipe network correlation of stations]{\label{Location-RC-A}\includegraphics[width=0.218\textwidth]{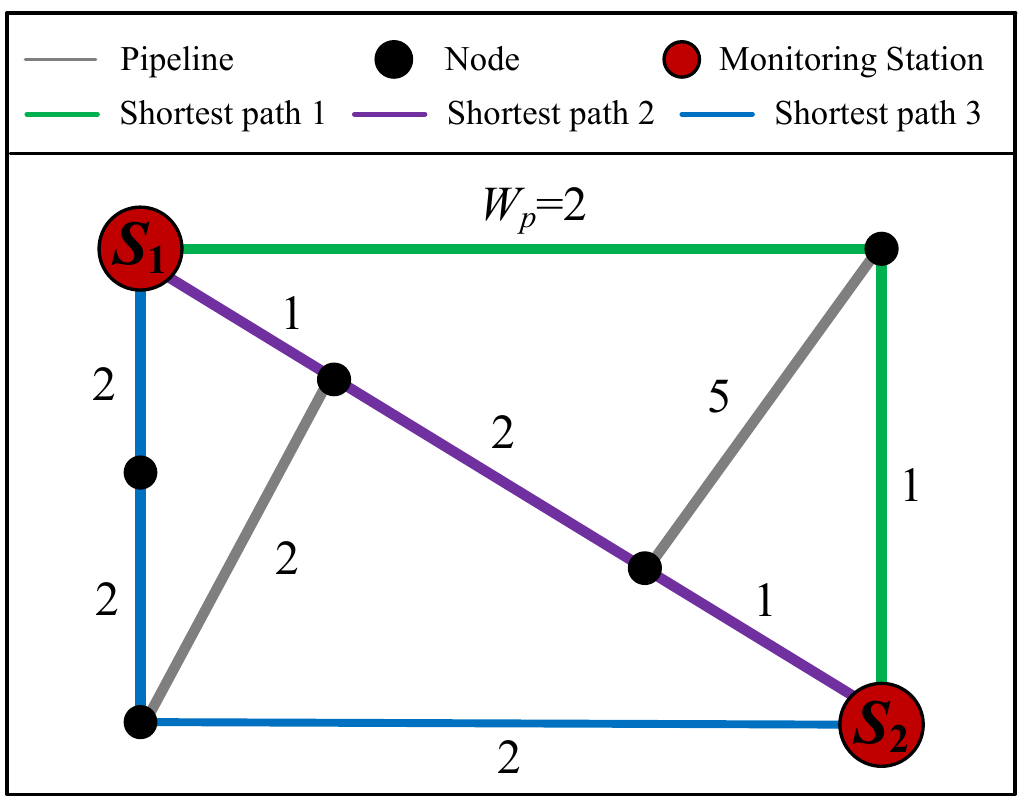}}
	\subfigure[RC readings of stations]{\label{Location-RC-B}\includegraphics[width=0.255\textwidth]{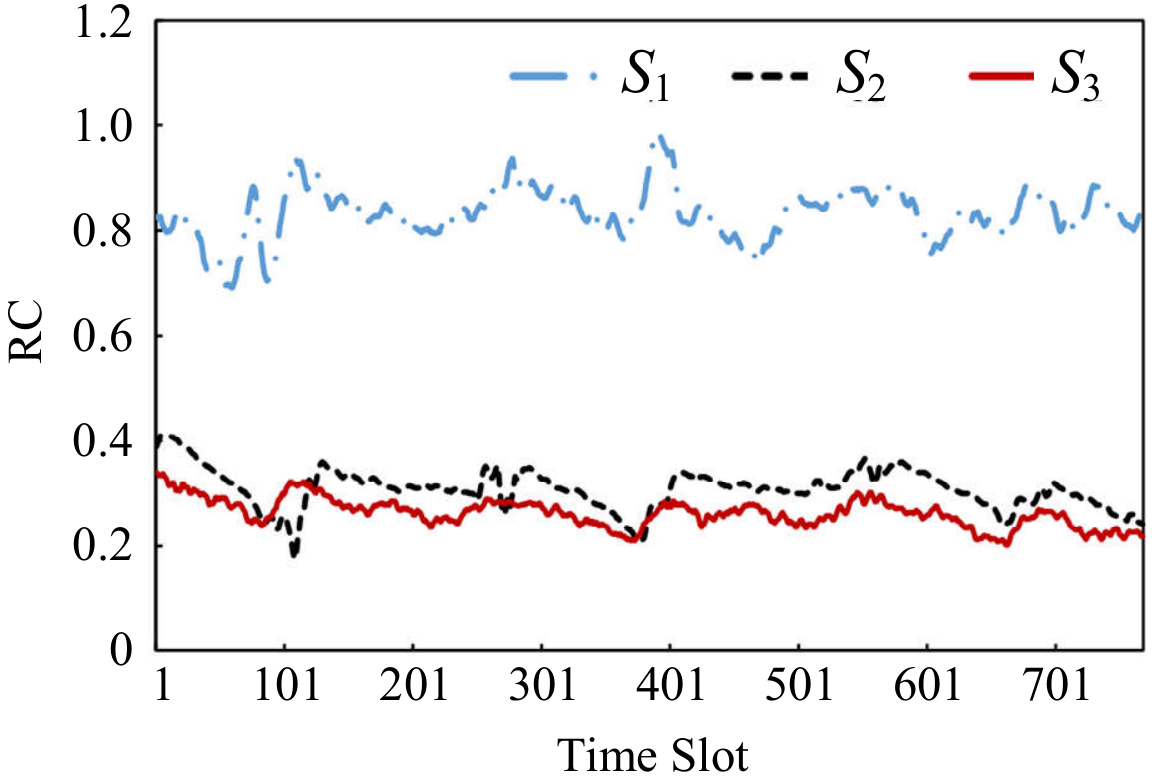}}
	\caption{Illustration of pipe network correlation among stations.}\label{Location-RC}
\end{figure}

Take the correlation between station $S_{1}$ and $S_{2}$ as an example, there are multiple paths between $S_{1}$ and $S_{2}$ in the pipe network. Given all of these paths, the correlation between $S_{1}$ and $S_{2}$ is affected by the pipe attributes along those paths. Specifically, this correlation is not only affected by the length of a pipe , but also influenced by other factors, including the diameter and age of a pipe. For each pipe $p$ in the network, we employ the three most important attributes (\textit{length}, \textit{diameter} and \textit{age}) and define its weight $w_{p}$ in this network as a function of the three attributes, i.e., $w_{p} = f(p.d, p.len, p.age)$ (we will give the definition of $w_{p}$ in later section). Figure \ref{Location-RC-A} illustrates the computation of correlation between $S_{1}$ and $S_{2}$, where the number along each edge denotes its weight $w_{p}$. As the interconnections between $S_{1}$ and $S_{2}$ are quite complicated, therefore a single path can hardly characterize their correlation; Consequently, we employ the top-k shortest paths as the characterization of those interconnections and calculate $C_{i, j}$ by the sum of top-k shortest paths with normalization. Specifically, $C_{i, j}$ is computed as follows:
\begin{equation}\label{CorrelationComputationEq}
	 C_{i, j} = \frac{1}{k} \sum_{t = 1}^{k}\sum_{p\in path(i, j, t)} w_{p},
\end{equation} 
where $path(i, j, t)$ represents the $t$-$th$ shortest path between $S_{i}$ and $S_{j}$. As shown in Figure \ref{Location-RC-A}, there are multiple paths between $S_{1}$ and $S_{2}$. When $k = 1$, we can get $C_{1, 2}$ by the shortest path (the green path) between $S_{1}$ and $S_{2}$, where $C_{1, 2} = \frac{1}{1} \sum_{t = 1}^{1}\sum_{p\in path(1, 2, t)} w_{p} = 2+1 = 3$. If we set $k = 3$, the correlation between $S_{1}$ and $S_{2}$ is computed by $C_{1, 2} = \frac{1}{3} \sum_{t = 1}^{3}\sum_{p\in path(1, 2, t)} w_{p} = \frac{(2+1) + (1+2+1) + (2+2+2)}{3} = \frac{13}{3}$.

Similarly, we also compute the correlation between $S_{2}$ and $S_{3}$ via Eqn. (\ref{CorrelationComputationEq}). In Figure \ref{Location-RC-B}, as compared to the correlation between $S_{1}$ and $S_{2}$, $S_{2}$ and $S_{3}$ has a larger correlation in the pipe network. From Figure \ref{Location-RC-B}, it can be seen that the concentrations of RC in $S_{2}$ and $S_{3}$ demonstrate the similar patterns, while the patterns and the readings in $S_{1}$ are quite different from $S_{2}$ and $S_{3}$, which verifies the existence of pipe network correlations among stations. Hence, capturing these spatial correlations can help predict the water quality accurately. One possible way of getting these spatial correlations is to use the Pearson correlation of the stations' historical data. However, this method has its own limitations if the historical data in some of stations is not available. Therefore, we take steps to another direction, which analyzes the spatial correlations by using the pipe network attributes.

\begin{figure}[!th]
	\centering
	\subfigure[Power of diameter]{\label{PowerOfDiameter}\includegraphics[width=0.158\textwidth]{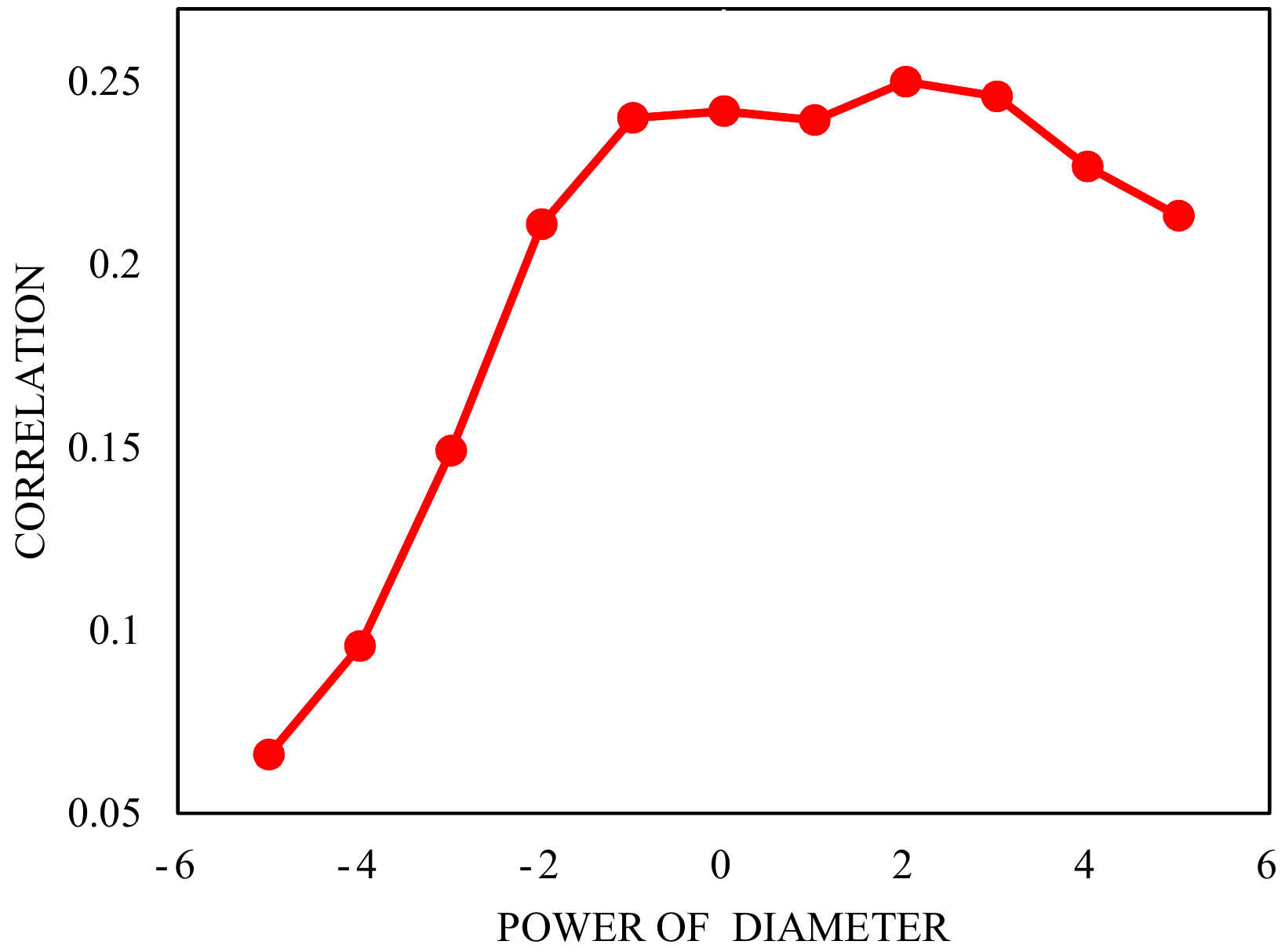}}
	\subfigure[Power of age]{\label{PowerOfAge}\includegraphics[width=0.158\textwidth]{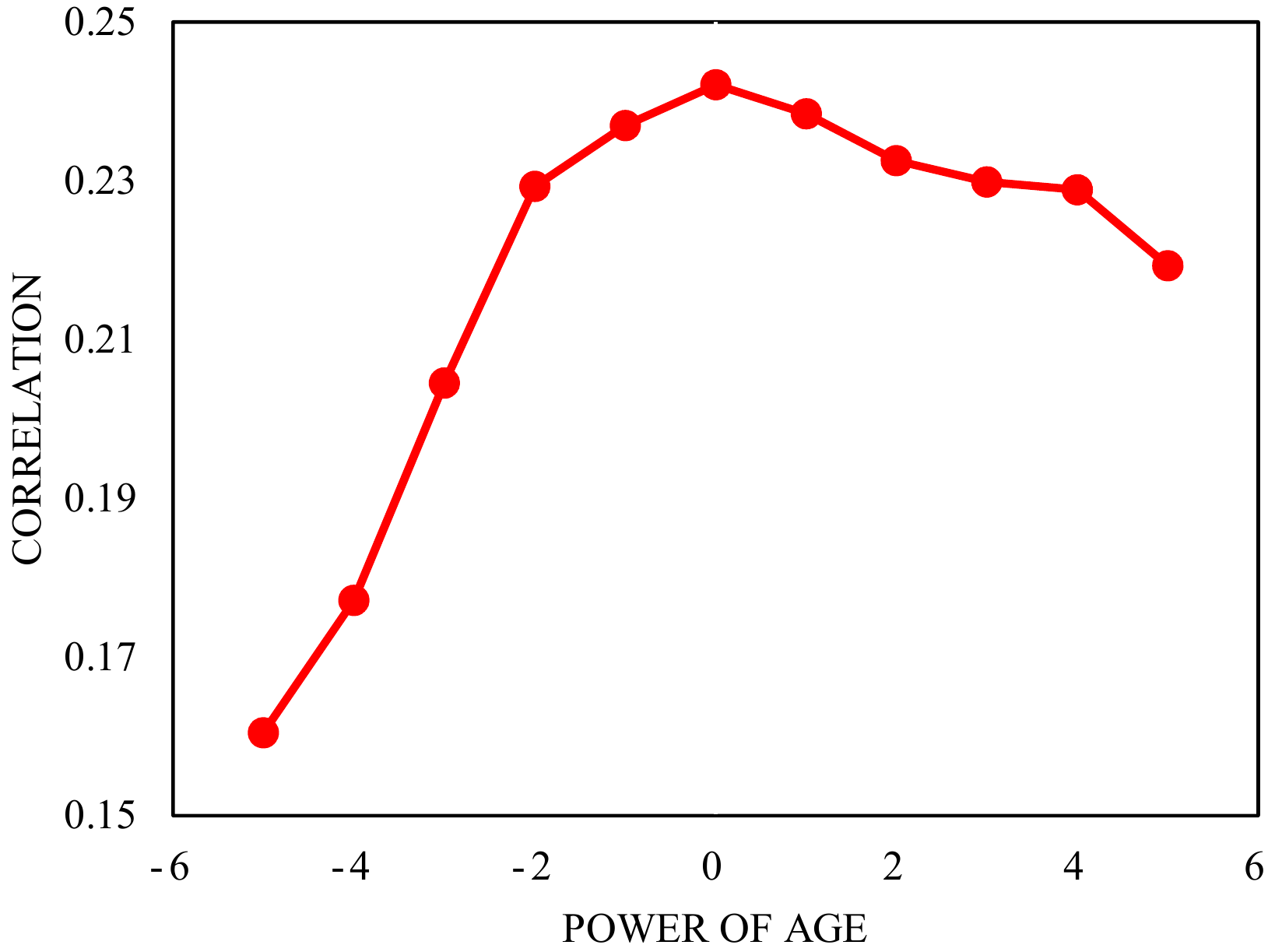}}
	\subfigure[Power of length]{\label{PowerOfLength}\includegraphics[width=0.158\textwidth]{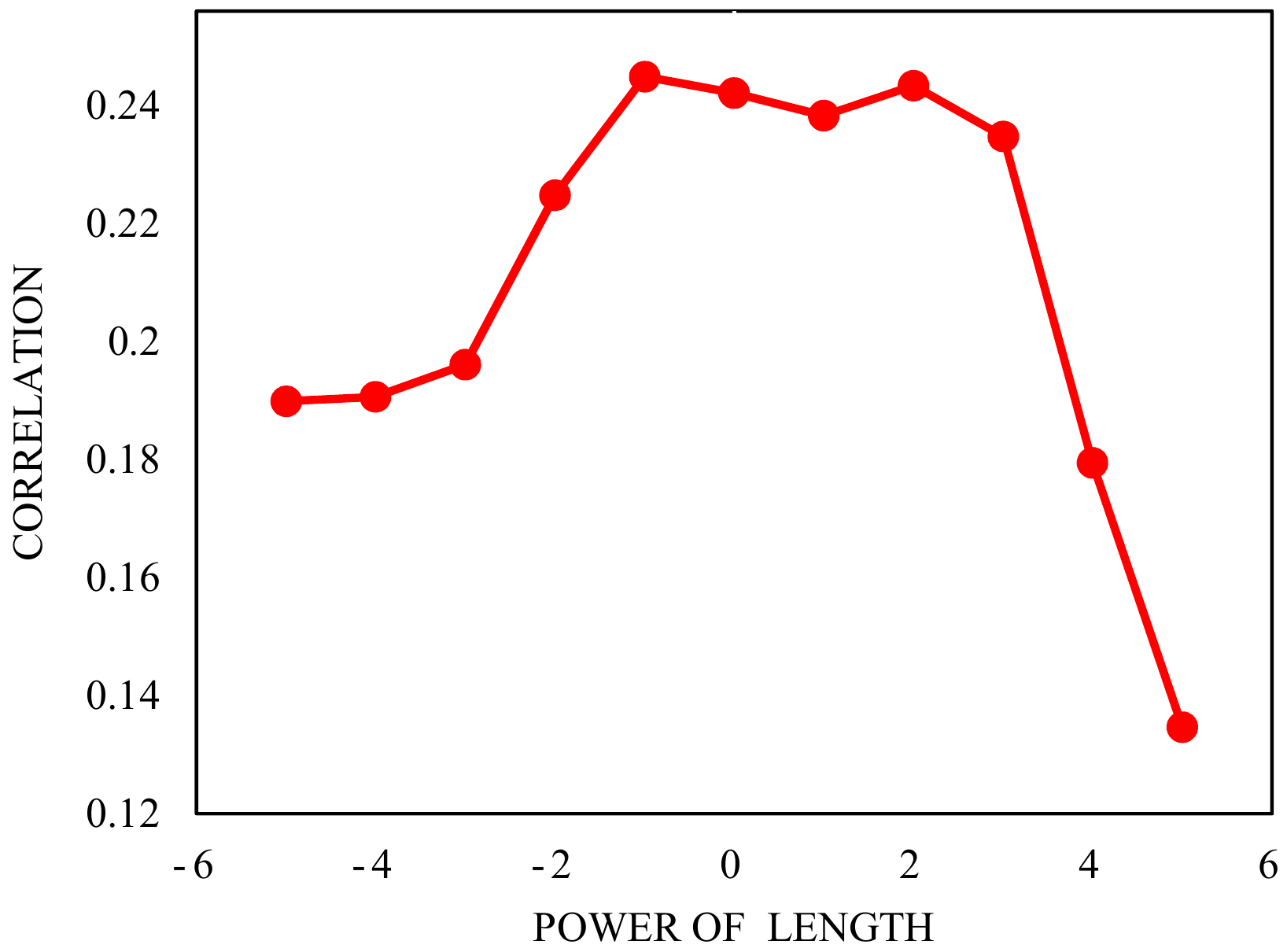}}
	\caption{Illustration of the impacts of various pipe attributes (\textit{diameter}, \textit{age}, \textit{length}) on the station correlations.}\label{PipeAttributesEffects}
\end{figure}

To get a comprehensive understanding of the effects, we investigate how different configurations of \textit{length}, \textit{diameter}, \textit{age} of a pipe can affect the correlations among different stations. Figure \ref{PipeAttributesEffects} comparatively illustrates the correlation results of all stations with respect to various pipe attributes by controlling other factors. In particular, we control the age and length by setting them to zero, and observe the effects of the power of diameter on the station correlations. From Figure \ref{PowerOfDiameter}, it can be observed that the correlation shows an increasing trend from $-5$ to $-1$ and stablize from $-1$ to $1$, then arrives its peek at $2$ followed by a decline. This demonstrate that, when keeping the other two factors (\textit{age}, \textit{length}) as zero, the square of diameter affects station correlations the most. Similarly, we can get the observations that age to the zero power and length to the negative one power have the greatest influence on the correlations while keeping the power of the other two as zeros.

Moreover, the optimal value of power for the \textit{diameter}, \textit{age} and \textit{length} of a pipe is determined by experiments. Table \ref{PipeAttributesCorrelations} illustrates the top power triplets < $pow\_d, pow\_len, pow\_age$ > which give the highest station correlations. In particular, the steps for getting the station correlations are as follows:
\begin{enumerate}[1)]
	\item Suppose there are $M$ stations in total, we can thus get a $M\times M$ correlation matrix $CorrMat$, which encodes the pairwise Pearson correlations of one year's RC readings from the $M$ stations.
	\item The pipe network can be seen as a weighted graph, where the weight $w$ for a pipe $p$ is computed from its attributes via $w_{p} = p.d^{pow\_d}\times p.len^{pow\_len} \times p.age^{pow\_age}$
	\item For each power triplet < $pow\_d, pow\_len, pow\_age$ >, where each entry in the triplet is iterated from $-5$ to $5$, we can get a $M\times M$ weight matrix $WeightMat$ by computing the top-$k$ shortest paths between any two stations.
	\item Thereafter, the correlation value in Table \ref{PipeAttributesCorrelations} is calculated by the correlation between $CorrMat$ and $WeightMat$, which indicates the degree of approximation of $WeightMat$ for the $CorrMat$.
\end{enumerate}

Interestingly, it can be observed that the power triplet < 2, -1, -1 >, which gives the second largest correlation result, is closest to the previous observations, i.e., the optimal value with the others fixed. Also, its correlation result is close to the top correlation result. Therefore, we choose the power triplet < 2, -1, -1 > as the power of \textit{diameter}, \textit{length} and \textit{age} for computing the weight $w_{p}$ of a pipe $p$. Specifically, $w_{p}$ is calculated by Eqn. $\ref{PipeWeightEquation}$,  

\begin{equation}\label{PipeWeightEquation}
	w_{p} = p.d^{2}\times p.len^{-1} \times p.age^{-1}.
\end{equation}

\begin{table}[!h]
	\begin{center}
		\caption{Illustration of correlations from the top power triplets.}\label{PipeAttributesCorrelations}
		\renewcommand{\arraystretch}{1.45}
		\begin{tabular}{|c|c|c|c|}
			\hline
			\begin{tabular}[c]{@{}c@{}}Power of \\ Diameter\end{tabular}  &  \begin{tabular}[c]{@{}c@{}}Power of \\ Length\end{tabular}  & \begin{tabular}[c]{@{}c@{}}Power of \\ Age\end{tabular}  &  Correlations   \\
			\hline
			\hline
			 1	 &  2   &  2  &  0.370383 \\
			\hline
			 \textbf{2}	 &  \textbf{-1}  & 	\textbf{-1}	&  \textbf{0.370336}\\
			\hline
			 0  &  2   &   2  &  0.367362\\
			\hline
			 2 &  -2   & 2 &  0.366843\\
			\hline
			 2 &  -1  & 0 &  0.366589\\
			\hline
			 2 &  2   & 2 &  0.365698\\
			\hline
			 0 &  2   & 1 &  0.364624\\
			\hline
		\end{tabular}
	\end{center}
\end{table}

To utilize these spatial correlations for co-predictions, we pre-compute those correlations through the structure of pipe network. In particular, the pipe network can be seen as a weighted graph, where the weight $w$ for a pipe $p$ is computed from the attributes of the pipe by $\frac{p.d^{2}}{p.len*p.age}$. Given station $S_{i}$ and $S_{j}$, as there are multiple different paths between $S_{i}$ and $S_{j}$, their pipe network correlation $C_{i, j}$ is computed by the sum of top-k shortest path between them.

\subsubsection{Technical Formulation}
Next, we will get into the technical details of the multi-task learning. Given the $M$ nodes for water quality prediction, the whole feature matrix $ \textbf{X}_{l} = [\textbf{X}_{l}^{s}, \textbf{X}_{l}^{t}] \in \mathbb{R}^{N_{l}\times D}$, and the target vector at $ \textbf{y}_{l} = \{\textbf{y}_{l,1}, \textbf{y}_{l,2}, \ldots, \textbf{y}_{l,N_{l}}\}\in \mathbb{R}^{N_{l}}$, for each node $l$, we can formulate the water quality prediction over all stations as a multi-task learning problem by considering the global impact on node $l$. In particular, we expand the model in Eqn. (\ref{sourceConsistence}) to incorporate a graph Laplacian penalty among node $l$ and the other nodes. This penalty ensures a small deviation between two nodes that are near in the pipeline system, and incorporates the domain knowledge about the spatial correlations of the water quality among different nodes in the pipeline systems. We formulate the multi-task learning problem as~\cite{belkin2006manifold_laplacian}\cite{ando2007learning_Laplacian}\cite{zhou2016GSpartan_MTL}
\begin{equation}\label{MTLPredictionObjectiveFunction}
\nonumber
   \mathop{\min}_{\textbf{W}}  \frac{1}{2}\sum_{l = 1}^{M}\|\textbf{y}_{l}-\frac{1}{2}\textbf{X}_{l}\textbf{w}_{l}\|_{2}^{2} + {\gamma}\sum_{l,m = 1}^{M}C_{l,m}\|\textbf{w}_{l} - \textbf{w}_{m} \|_{2}^{2},
\end{equation}
where $\gamma$ is a regularization parameter, $C_{l,m}$ is the pipe network correlation between task (station) $l$ and task (station) $m$, and measures the spatial autocorrelation between task $l$ and $m$. Intuitively, if $ C_{l,m}$ is large, the graph Laplacian regularizer term will force $\textbf{w}_{l}$ to be as similar as $\textbf{w}_{m}$. Thus, this graph Laplacian penalty automatically encodes Tobler's first law of geography~\cite{tobler1970computer}. In implementation, we can pre-compute the correlation $C_{l,m}$ by top-k shortest path in the pipe networks.

\subsection{Optimization}
The optimization of Eqn.$(\ref{PredictionObjectiveFunction})$ is convex with respect to $\textbf{W}$. First, we can rewrite the graph Laplacian term in Eqn.$(\ref{PredictionObjectiveFunction})$ as
\begin{equation}\label{GraphLaplacian}
   \sum_{l,m = 1}^{M}C_{l,m}\|\textbf{w}_{l} - \textbf{w}_{m} \|_{2}^{2} = tr(\textbf{W}(\textbf{D}-\textbf{C})\textbf{W}^{T}) = tr(\textbf{W}\textbf{L}\textbf{W}^{T})
\end{equation}
where $\textbf{D}$ is a diagonal matrix with $D_{l,l} = \sum_{m}C_{l,m}$, $\textbf{C}$ is the similarity matrix, and $\textbf{L} = \textbf{D}-\textbf{C}$ is known as the Laplacian matrix~\cite{belkin2006manifold_laplacian}\cite{ando2007learning_Laplacian}. We define
\begin{eqnarray}
  \nonumber
  h(\textbf{W}) &=& \frac{1}{2}\sum_{l = 1}^{M}\|\textbf{y}_{l} - \frac{1}{2}\textbf{X}_{l}\textbf{w}_{l}\|_{2}^{2} + {\lambda}\sum_{l = 1}^{M}\|\textbf{X}_{l}^{s}\textbf{w}_{l}^{s}-\textbf{X}_{l}^{t}\textbf{w}_{l}^{t}\|_{2}^{2} \\
  &&+ {\gamma}tr(\textbf{W}\textbf{L}\textbf{W}^{T}), \\
  g(\textbf{W}) &=& {\theta}\| \textbf{W} \|_{2,1}.
\end{eqnarray}

The optimization in Eqn.$(\ref{PredictionObjectiveFunction})$ can be rewritten as $\small\mathop{\min}_{\textbf{W}} h(\textbf{W}) + g(\textbf{W})$, where $\small h(\textbf{W})$ is smooth and $\small g(\textbf{W})$ is non-smooth. We can thus use the Fast Iterative Shrinkage-Thresholding Algorithm (FISTA) ~\cite{beck_SIAM2009_FASTA} or Accelerated Gradient Descent (AGD)~\cite{zhou2011malsar} to solve it. One of the key steps in using FISTA algorithm is to solve the proximal step~\cite{beck_SIAM2009_FASTA}:
\begin{equation}\label{FISTA_derivations}
\small
  \textbf{W}^{(k)} = \argmin_{\textbf{W}}\{g(\textbf{W}) + \frac{L_{k}}{2}\| \textbf{W} - (\textbf{V}^{(k)} - \frac{1}{L_{k}}\nabla h(\textbf{V}^{(k)})) \|^{2}_{F} \},
\end{equation}
where $\small \textbf{V}^{(k)}$ is the search point and is defined by the affine combination of $\small \textbf{W}^{(k-1)}$ and $\small \textbf{W}^{(k-2)}$; and $\small L_{k}$ is a scalar that can be determined by the line search method~\cite{beck_SIAM2009_FASTA}. The gradient $\small \nabla h(\textbf{W})$ for each column $\small\textbf{w}_{l}$ in Eqn. (\ref{FISTA_derivations}) can be computed as
\begin{equation}\label{FISTA_gradient}
\small
  \nabla h(\textbf{w}_{l}) = \frac{1}{2}\textbf{X}_{l}^{T}(\frac{1}{2}\textbf{X}_{l}\textbf{w}_{l}-\textbf{y}_{l}) + \lambda\textbf{P}\textbf{w}_{l} + 2\gamma\textbf{W}\textbf{L}_{l},
\end{equation}
where $\small\textbf{L}_{l}$ is the $\small l$-$\small th$ column of $\small\textbf{L}$, $\small \textbf{P}\in \mathbb{R}^{D\times D}$ is a block matrix with $\small 2\times 2$ blocks, and $\small \textbf{P}_{11} = 2{(\textbf{X}_{l}^{s})}^{T}\textbf{X}_{l}^{s}$, $\small \textbf{P}_{12} = -2{(\textbf{X}_{l}^{s})}^{T}\textbf{X}_{l}^{t}$, $\small \textbf{P}_{21} = -2{(\textbf{X}_{l}^{t})}^{T}\textbf{X}_{l}^{s}$, $\small \textbf{P}_{22} = 2{(\textbf{X}_{l}^{t})}^{T}\textbf{X}_{l}^{t}$.

As Eqn.$(\ref{FISTA_derivations})$ is computed in every FISTA iteration, it needs to be solved efficiently.
Specifically, each row $\textbf{w}^{i}$ of $\small\textbf{W}$ can be decoupled in Eqn.$(\ref{FISTA_derivations})$. Thus, to obtain the $i$-$th$ row $\small\textbf{w}^{i(k)}$ in $\small\textbf{W}^{(k)}$, we only need to solve the following problem:
\begin{equation}\label{update_RowofW}
\small
  \textbf{w}^{i(k)} = \argmin_{\textbf{w}}\{\theta \|\textbf{w}\|_{2} + \frac{L_{k}}{2}\| \textbf{w} - (\textbf{v}^{i(k)} - \frac{1}{L_{k}}\nabla f(\textbf{v}^{i(k)})) \|^{2}_{2} \},
\end{equation}
where $\textbf{v}^{i(k)}$ is the $i$-$th$ row of $\textbf{V}^{(k)}$. It can be reformulated as
\begin{equation}\label{update_w_i}
\small
  \textbf{w}^{i(k)} = \argmin_{\textbf{w}}\{ \frac{1}{2}\|\textbf{w} - \textbf{b} \|^{2}_{2} + \beta\| \textbf{w} \|_{2} \},
\end{equation}
where $\small\textbf{b} = \textbf{v}^{i(k)} - \frac{1}{L_{k}}\nabla f(\textbf{v}^{i(k)})$ and $\small\beta = \frac{\theta}{L_{k}}$. It has been proved by Wright \emph{et al.}~\cite{WrightSpaRSA_2009} that Eqn.$(\ref{update_w_i})$ have the closed-form solution,
\begin{equation}\label{closeform_w_i}
\small
  \textbf{w}^{i(k)} = max(0,1-\frac{\beta}{\|\textbf{b}\|_2})\textbf{b}.
\end{equation}

We thus can solve Eqn.$(\ref{FISTA_derivations})$ quite efficiently. The FISTA algorithm is depicted in Algorithm \ref{FISTA_algorithm}, where $\|\textbf{W}\|_{F}$ is the Frobenius norm and $\small L_{k}$ is a scalar that can be determined by the line search method~\cite{beck_SIAM2009_FASTA}.

\begin{algorithm}
\caption{FISTA algorithm for solving Eqn.$(\ref{PredictionObjectiveFunction})$}\label{FISTA_algorithm}
\begin{algorithmic}[1]
\STATE \textbf{Input:} Iterations $K$
\STATE \textbf{Initialize:} $\textbf{V}^{(1)} = \textbf{W}^{(0)} \in \mathbb{R}^{D\times M}, t_{1} = 1$\\
\FOR{$k = 1, 2, 3, \ldots, K$}
    \STATE $\textbf{W}^{(k)} \leftarrow \argmin_{\textbf{W}}\{g(\textbf{W}) + \frac{L_{k}}{2}\| \textbf{W} - (\textbf{V}^{(k)} - \frac{1}{L_{k}}\nabla f(\textbf{V}^{(k)})) \|^{2}_{F} \}$
    \STATE $t_{k+1} \leftarrow \frac{1 + \sqrt{1 + 4t_{k}^{2}}}{2}$
    \STATE $\textbf{V}^{(k+1)} \leftarrow \textbf{W}^{(k)} + (\frac{t_{k} - 1}{t_{k+1}})(\textbf{W}^{(k)} - \textbf{W}^{(k-1)})$
\ENDFOR
\STATE \textbf{Output:} $\textbf{W}^{(K)}$
\end{algorithmic}
\end{algorithm}

\subsection{Computational Complexity Analysis}
In this section, we discuss the computational complexity for solving the stMTMV model. For the optimization of $\small \textbf{W}$, the main computational cost comes from calculating the gradient $\small \nabla h(\textbf{W})$ in each FISTA iteration. In particular, as $\small \textbf{X}_{l}^{T}\textbf{X}_{l}$, $\small \textbf{X}_{l}^{T}\textbf{y}_{l}$ and $\small \textbf{P}$ remain unchanged in every FISTA iteration, we can pre-compute and store them with time complexity of $\small O(ND^{2})$. In each FISTA iteration, the time complexity for calculating the gradient $\small \nabla h(\textbf{W})$ in Eqn. (\ref{FISTA_gradient}) is $\small O((D+M)DM)$, and computing the closed-form solution in Eqn.$(\ref{closeform_w_i})$ takes $\small O(DM)$ time for $\small \textbf{W}^{(k)}$. Hence, the complexity for each iteration in the FISTA algorithm is $\small O((D+M)DM)$. Moreover, the FISTA algorithm converges within $O(1/\epsilon^{2})$ iterations, and the total time cost of FISTA for solving stMTMV is $\small O(\frac{(D+M)DM}{\epsilon^2})$, where $\epsilon$ is the desired accuracy. Thus, the stMTMV model can be solved efficiently. Since the per-iteration complexity of FISTA for solving stMTMV is independent of $N$, which shows that our model can potentially scale to large-scale urban data.

\section{Evaluation}\label{Experiments}
\subsection{Datasets}
In the evaluation, we use the following six real datasets detailed in Table \ref{DatasetStatistics}, where the data are collected from August 2011 to August 2014 in Shenzhen City, China.

\begin{table}[!h]
\begin{center}
\caption{Details of the datasets}\label{DatasetStatistics}
\renewcommand{\arraystretch}{1.45}
\begin{tabular}{|c|c|c|}
  \hline
    \multicolumn{2}{|c|}{\textbf{Data Sources}}         & \textbf{Shenzhen}             \\ \cline{2-3}
  \hline
    \multirow{3}{*}{Water Quality}                      & Stations                      & 14  \\ \cline{2-3}
                                                        & Time Spans                    & 8/1/2011-7/31/2014  \\ \cline{2-3}
                                                        & Hours                         & 17,544  \\ \cline{2-3}
  \hline
    \multirow{2}{*}{POIs}                               & POIs                          & 185,841  \\ \cline{2-3}
                                                        & Categories                    & 20 \\ \cline{2-3} \hline
    \multirow{2}{*}{Road Network}                       & Segments                      & 149,559  \\ \cline{2-3}
                                                        & Total length                  & 15,132km  \\ \cline{2-3}
  \hline
    \multirow{3}{*}{Pipe Network}                       & Nodes                         & 8,707  \\ \cline{2-3}
                                                        & Segments                      & 6,152  \\ \cline{2-3}
                                                        & \begin{tabular}[c]{@{}c@{}}Max Connected\\Component\end{tabular}  & 3,511  \\ \cline{2-3} \hline
                      Meteorology                       & Hours                         & 55,633  \\ \cline{2-3} \hline
    \multirow{3}{*}{Hydraulic Data}                     & Flow Stations                 & 10  \\ \cline{2-3}
                                                        & Pressure Stations             & 13  \\ \cline{2-3}
                                                        & Time Spans                    & 8/1/2011-7/31/2014  \\ \cline{2-3} \hline
\end{tabular}
\end{center}
\end{table}

\begin{enumerate}[1)]
  \item Water quality data: We collect water quality data every five minutes from 15 water quality monitoring stations in Shenzhen City. It comprises residual chlorine (RC), turbidity (TU) and pH. In this paper, we only use RC as the index for water quality, since RC is the most important and effective measurement for water quality in current urban water distribution system~\cite{rossman_RCmodeling_1994}.
  \item Hydraulic data: Hydraulic data consists of flow and pressure, which are collected every five minutes from 13 flow sites and 14 pressure sites, respectively.
  \item Road networks data: Each road segment is associated with two terminal points and some properties, such as level, capacity and speed limit.
  \item Pipe attributes data: It describes the pipe attributes in water distribution system, and the attributes of a pipe consist of diameter, length, age, material, etc.
  \item Meteorology data: Meteorological data consists of weather, temperature, humidity, barometer pressure, wind strength, which is collected every hour.
  \item POIs: There are 185841 POIs of 20 categories. Each POI has a name, category, address and geo-coordinates.
\end{enumerate}

\subsection{Ground Truth and Metrics}
We can predict the water quality of a site from its historical data, and the ground truth is obtained from its later readings. In particular, we evaluate the predictive performance with respect to its readings in next $1$, $2$, $3$, $4$ hours, and the performance is evaluated in terms of their root-mean-square-error (RMSE):
\begin{equation}\label{RMSE}
    RMSE = \sqrt{\frac{1}{N}\sum_{l = 1}^{M}(\textbf{y}_{l}-\hat{\textbf{y}}_{l})^{2}}.
\end{equation}

\subsection{Learning Model Comparison}\label{ModelComparisonSection}
To validate our stMTMV model, we compared it with the following baselines:
\begin{itemize}
  \item RC Decay Model (Classical): Residual Chlorine (RC) decay model is a simple (also well-known) mathematical model that is typically applied in environmental science to model and predict chlorine residual in water supply systems~\cite{monteiro2014modeling}\cite{rossman1996numerical}\cite{rossman_RCmodeling_1994}. This model describes both bulk and wall chlorine consumption via first order decay kinetics $\frac{dC}{dt} = -kC$, where $k$ is the first order chlorine decay constant that depends on the distribution systems.
  \item ARMA: Auto-Regression-Moving-Average (ARMA) is a well-known model for predicting time series data, which makes predictions solely based on historical data~\cite{fabozzi2014autoregressive_ARMA}.
  \item Kalman: Kalman filter is a traditional predictive model that uses a series of measurements observed data over time and predicts the future value in a probabilistic manner~\cite{harvey1990forecasting}.
  \item ANN: Artificial Neural Network (ANN) is feeded with all features from different views into a single model without treating different views differently. Employing ANN as a baseline is to justify the advantages of using a combination of multiple views. As ANN has also been used in environmental science to predict water quality~\cite{kalin2010prediction}\cite{palani2008ann}, surpassing ANN also justifies our contribution over traditional approaches.
  \item LR: Linear Regression (LR)~\cite{seber2012linear_LR} is applied for each node individually, which is a single-task learning method.
  \item LASSO: \textbf{Lasso}~\cite{tibshirani_LASSO1996regression} tries to minimize the objective function $ \frac{1}{2}\sum_{l = 1}^{M}\|\textbf{y}_{l}-\textbf{X}_{l}\textbf{w}_{l}\|_{2}^{2} + \alpha\| \textbf{W} \|_{1}$ and encodes the sparsity over all weights in $\textbf{W}$. It keeps task-specific features but ignores the task-sharing features.
  \item MRMTL: As a typical example of traditional multi-task learning, Mean-Regularized Multi-Task Learning (MRMTL)~\cite{evgeniou2004regularized}\cite{zhou2011malsar} assumes all tasks are related and penalizes the deviation of each task from their mean by optimizing $\frac{1}{2}\sum_{l=1}^{M}\|\textbf{y}_{l}-\textbf{X}_{l}\textbf{w}_{l}\|_{2}^{2} + \lambda \sum_{l=1}^{M}\| \textbf{w}_{l} - \frac{1}{M}\sum_{m=1}^{M} \textbf{w}_{m} \|_{2}^{2} + \theta\| \textbf{W} \|_{F}^{2}$.
  \item regMVMT: The regularized multi-view multi-task learning model (regMVMT)~\cite{zhang2012inductive_regMVMT_KDD2012} jointly regularizes view consistency and uniform task \mbox{relatedness}.
  \item stMTMV-\textit{IJCAI}: Our previous results in IJCAI-16~\cite{MVMTL_IJCAI2016_UrbanWater}.
\end{itemize}

The experimental results are demonstrated in Table \ref{ModelComparison}. From this table, we have the following observations: 1) The prediction accuracy of all models shows a decrease trend for the next 1-4 hours. This is consistent with the intuition that the distant future tends to be more difficult to forecast than the near future. 2) The last six machine learning approaches demonstrate superior performance over first three methods, which justifies the effectiveness of extracted features as well as the benefits of data fusion from multiple sources. Moreover, it is not unexpected that RC Decay Model achieves the worst performance since it may fail to capture the real dynamics of the RC in the system. 3) The last four multi-task learning methods stably outperform the single-task learning methods (ANN, LR), which verifies that the tasks are not independent and that jointly learning them can boost learning performance. 3) The accuracies of MRMTL are slightly lower than other multi-task learning methods. This may be caused by the inappropriate assumption of penalizing the deviation of each task from their mean, since these tasks tend to be spatially autocorrelated. 4) As compared to MTL, our model and regMVMT achieve higher performance due to the fact that stMTMV and regMVMT can incorporate heterogeneous information from spatial and temporal views, which may help to improve overall performance. 5) The stMTMV model shows superiority over regMVMT, which underscores the importance of incorporating structure of the water distribution system and this structure can further improve performance. 6) The outperformance of stMTMV over stMTMV-\textit{IJCAI} demonstrates that the refined $C_{i, j}$ can better characterize the geographical correlation among stations and hence boost the overall performance as a consequence.

\begin{table}[!h]
\begin{center}
\caption{Predictive performance comparison among various approaches over next 1-4 hours.}\label{ModelComparison}
\renewcommand{\arraystretch}{1.45}
\begin{tabular}{|c||c|c|c|c|}
  \hline
  \small{Model Comparison}      & 1 hour  &  2 hour    & 3 hour  & 4 hour   \\
  \hline
  \hline
  RC Decay Model          & $3.51e$-$1$ &  $3.53e$-$1$   & $3.59e$-$1$ &  $3.68e$-$1$\\
  \hline
  ARMA                    & $1.86e$-$1$ &  $2.18e$-$1$   & $2.46e$-$1$ &  $2.78e$-$1$\\
  \hline
  Kalman                  & $1.79e$-$1$ &  $1.93e$-$1$   & $1.89e$-$1$ &  $1.90e$-$1$\\
  \hline
  ANN                     & $1.50e$-$1$ &  $1.54e$-$1$   & $1.56e$-$1$ &  $1.63e$-$1$\\
  \hline
  LR                      & $1.68e$-$1$ &  $1.99e$-$1$   & $2.09e$-$1$ &  $2.10e$-$1$\\
  \hline
  LASSO                   & $1.23e$-$1$ &  $1.42e$-$1$   & $1.52e$-$1$ &  $1.56e$-$1$\\
  \hline
  MRMTL                   & $1.32e$-$1$ &  $1.48e$-$1$   & $1.56e$-$1$ &  $1.58e$-$1$\\
  \hline
  regMVMT                 & $1.06e$-$1$ &  $1.15e$-$1$   & $1.18e$-$1$ &  $1.19e$-$1$\\
  \hline
  stMTMV-\textit{IJCAI}       & $9.33e$-$2$ &  $9.66e$-$2$   & $9.80e$-$2$ &  $9.90e$-$2$\\
  \hline
  stMTMV                  & $\textbf{8.99e}$-$\textbf{2}$ &  $\textbf{9.54e}$-$\textbf{2}$   & $\textbf{9.71e}$-$\textbf{2}$ &  $\textbf{9.83e}$-$\textbf{2}$\\
  \hline
\end{tabular}
\end{center}
\end{table}

\subsection{Evaluation on Model Components}
To evaluate each component of the stMTMV model, we compared it with three different variants of stMTMV:
\begin{itemize}
  \item stMTMV-\emph{us}: In this variant, uniform spatial correlation is used to evaluate the importance of spatial correlation among tasks. We can derive it by setting $\small \textbf{S} = \textbf{I}$.
  \item stMTMV-\emph{ws}: This is a derivation of stMTMV without group sparsity. We can derive it by setting $\small \theta = 0$.
  \item stMTMV-\emph{sv}: This derivation is to evaluate the importance of spatio-temporal view alignment. We can derive it by setting $\small \lambda = 0$.
\end{itemize}

\begin{figure}[!h]
  \centering
  \includegraphics[width=0.49\textwidth]{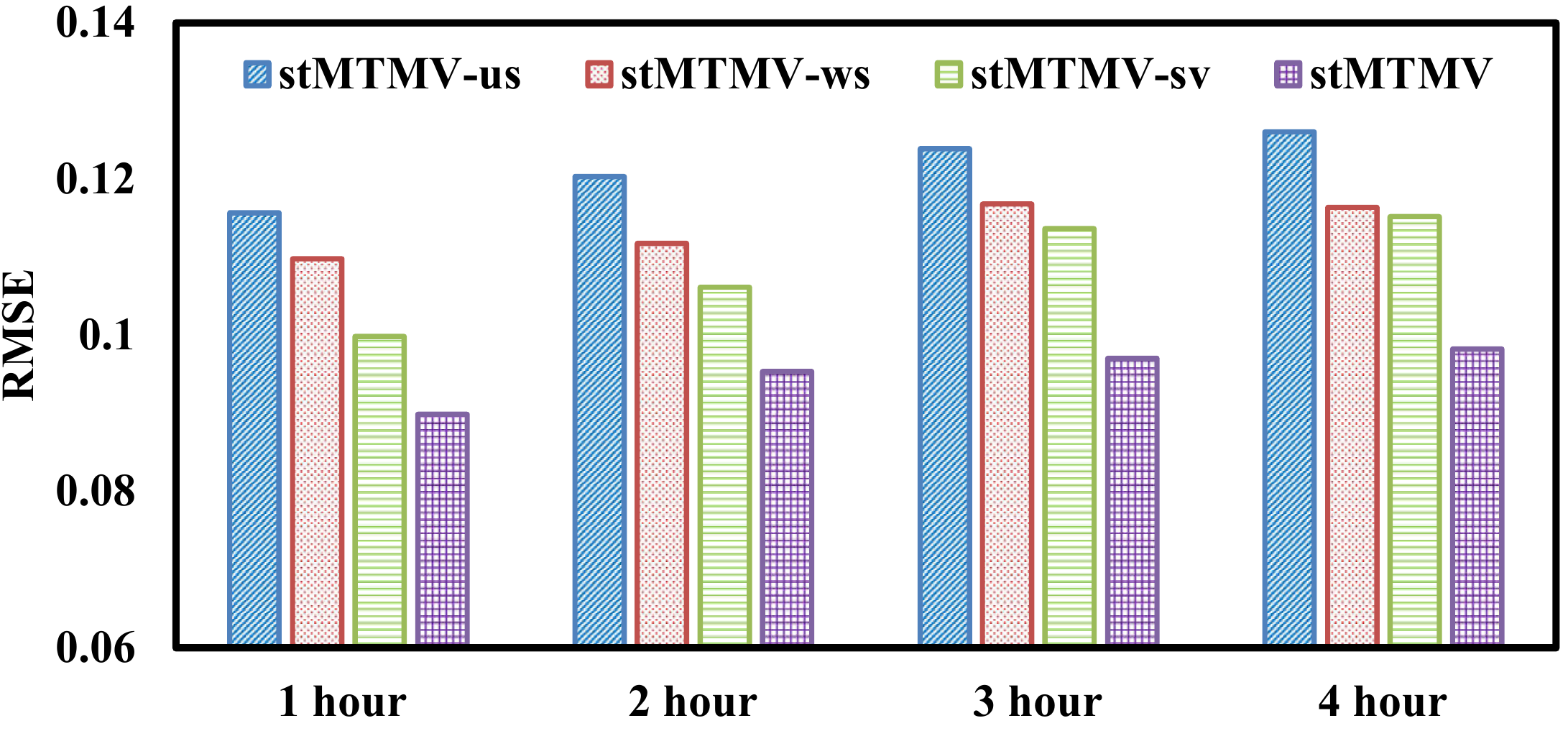}\\
  \caption{Comparative predictive performance illustration of each component in stMTMV with respect to 1-4 hours.}\label{ComponentComparisons}
\end{figure}

The experimental results are demonstrated in Figure \ref{ComponentComparisons}. From this figure, it can be seen that
stMTMV-\emph{us} achieves the worst performance, which demonstrates the effectiveness of graph Laplacian component in the stMTMV model. This further verifies that the tasks are mutually correlated and the spatial autocorrelation plays an important role in the co-prediction tasks. Moreover, stMTMV-\emph{ws} achieves the second worst performance, which justifies the importance of group sparsity in the stMTMV model. This also provides evidence for the assumption that only a small set of features are predictive for the water quality prediction tasks. Compared to stMTMV-\emph{ws} and stMTMV-\emph{us}, the effect of spatio-temporal view alignment tend to be less strong, and this is observed by the superior performance of stMTMV-\emph{sv} over other two variants. However, stMTMV outperforms stMTMV-\emph{sv} since spatio-temporal view alignment can combine heterogeneous spatio-temporal information and further boost performance.

\subsection{Evaluation on Views}
To demonstrate the descriptiveness of each view, we compared our stMTMV model over the following combinations.
\begin{itemize}
  \item \emph{t-view}: Only temporal view (\emph{t-view}) is used.
  \item \emph{s-view}: Only spatial-view (\emph{s-view}) is used.
  \item \emph{st-view-na}: Both spatio-temporal views are used, but there is no \emph{s-t view} alignment within each station.
  \item \emph{st-view}: Both spatio-temporal view are used and the \emph{s-t view} alignment is employed for each station.
\end{itemize}

\begin{figure}[!h]
  \centering
  \includegraphics[width=0.49\textwidth]{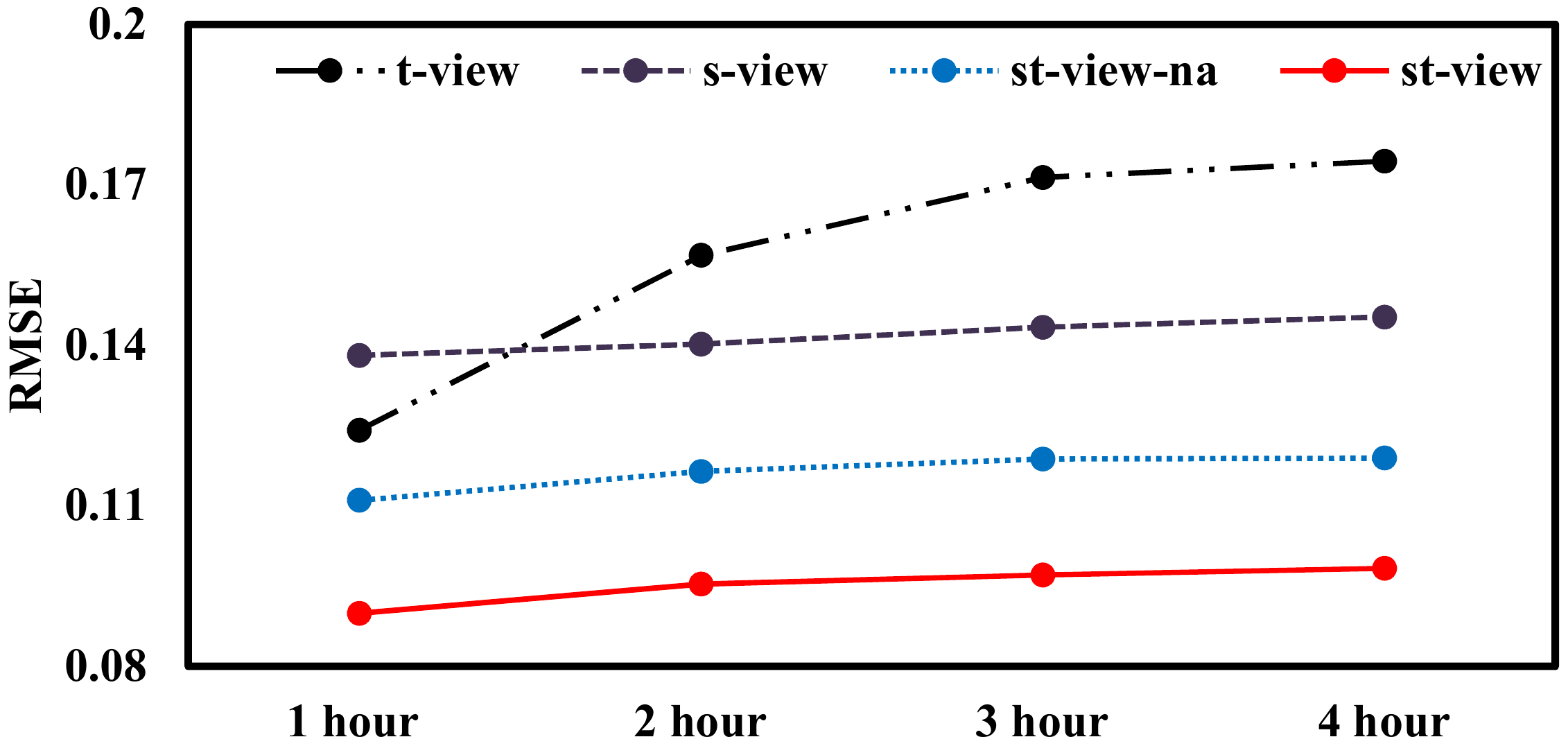}\\
  \caption{Comparative predictive performance illustration of stMTMV over view combinations with respect to 1-4 hours.}\label{ViewComparisons}
\end{figure}

The results are presented in Figure \ref{ViewComparisons}. From this figure, we observe that: 1) the combinations of spatial and temporal views outperform each individual one. This observation reveals that the more views fed to our model, the better the performance that can be achieved. 2) the \emph{st-view} outperforms \emph{st-view-na}, which implies that aggregating information from spatial and temporal views can achieve better performance than concatenating them together. This also verifies that the heterogeneous information distributed across spatial and temporal views is usually complementary rather than conflicting, and appropriate aggregation of these can provide a better way to capture each station's characteristics comprehensively, and consequently boost the performance.

\subsection{Predictive Performance among Stations}\label{StationComparison}
\begin{figure*}
  \centering
  \subfigure[Learning model evaluation]{\label{ModelComparisonNode}\includegraphics[width=0.333\textwidth]{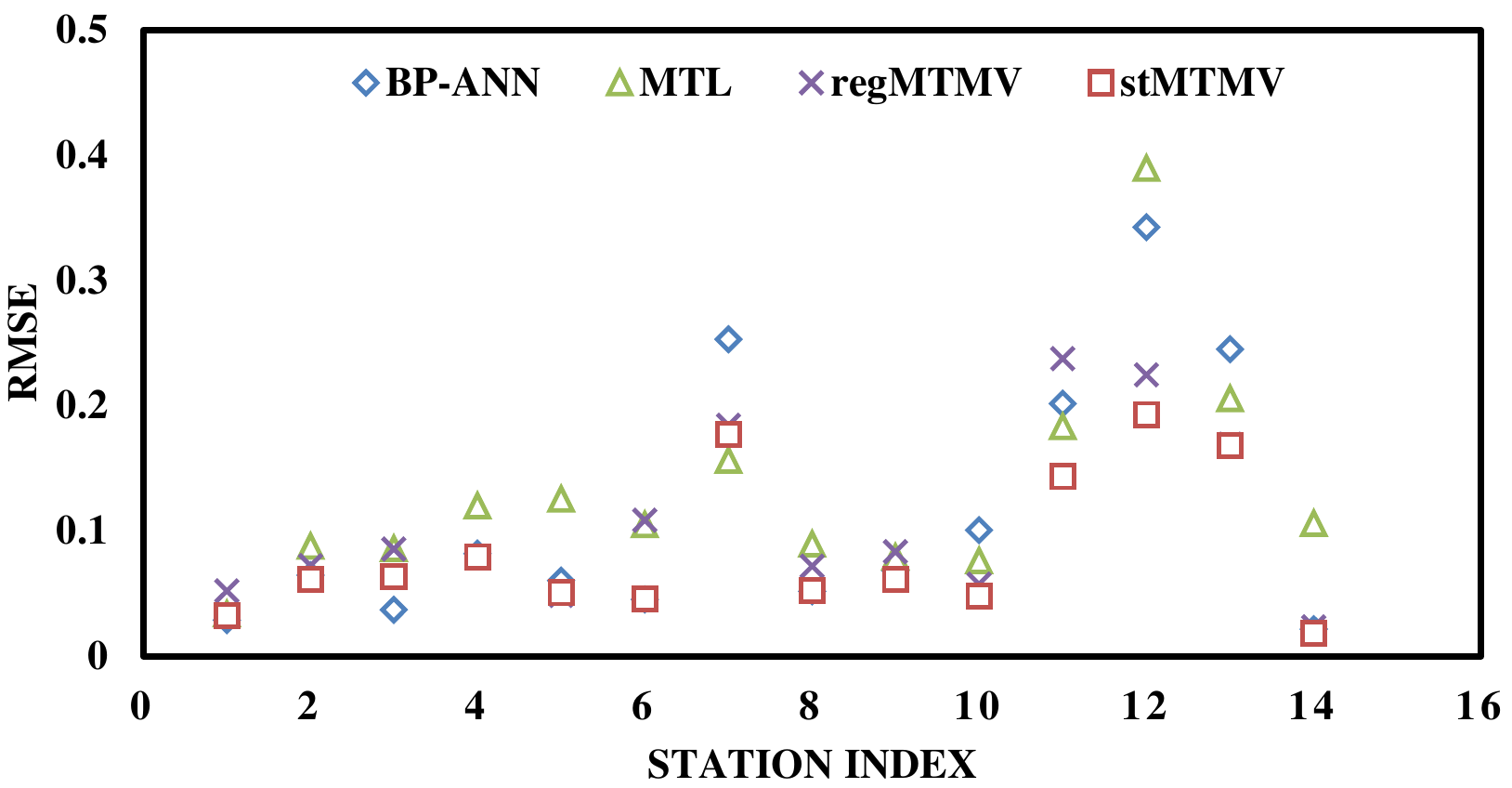}}
  \subfigure[Model variants evaluation]{\label{ModelComponentNode}\includegraphics[width=0.325\textwidth]{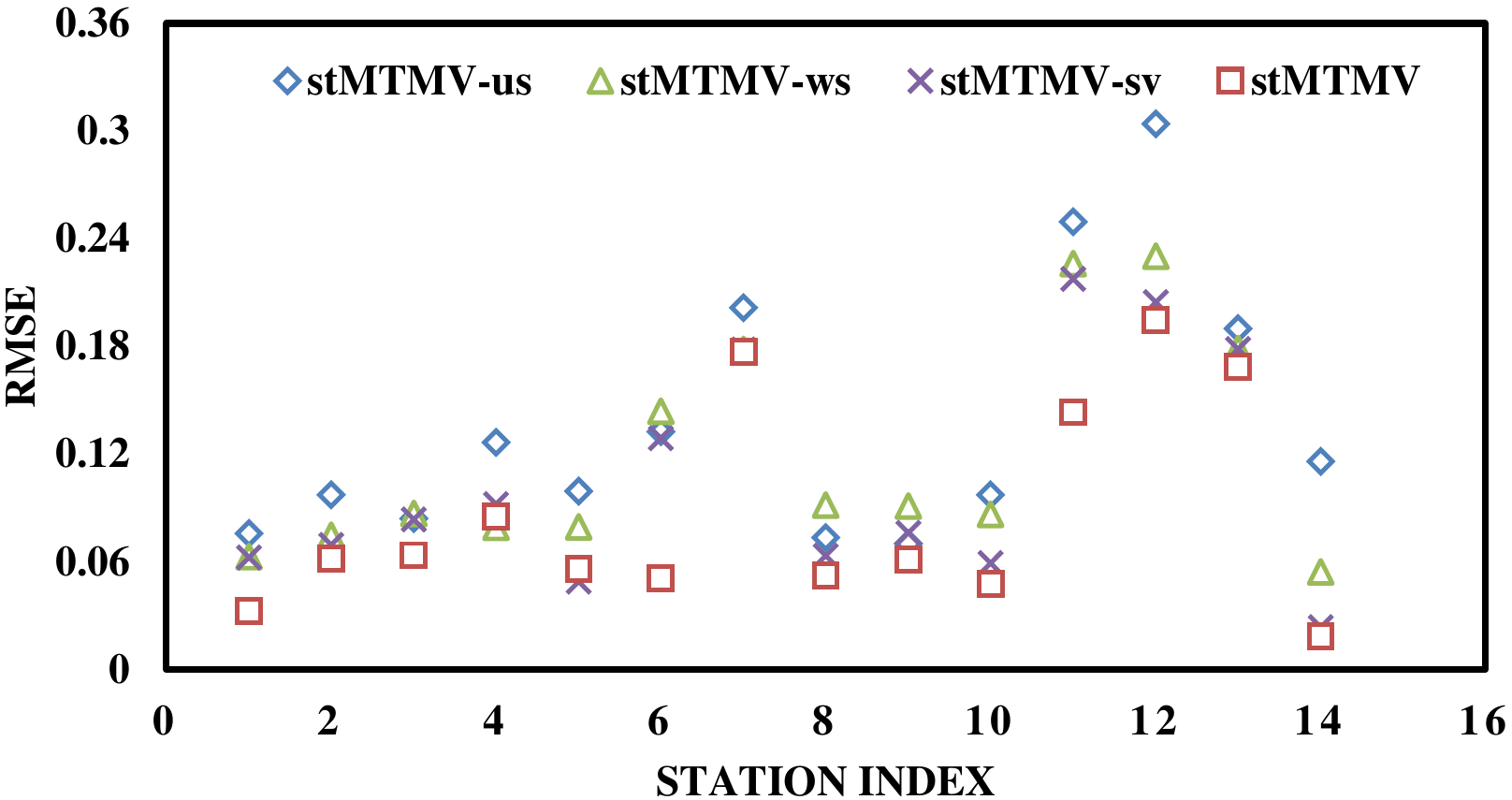}}
  \subfigure[View evaluation]{\label{ViewComparisonNode}\includegraphics[width=0.33\textwidth]{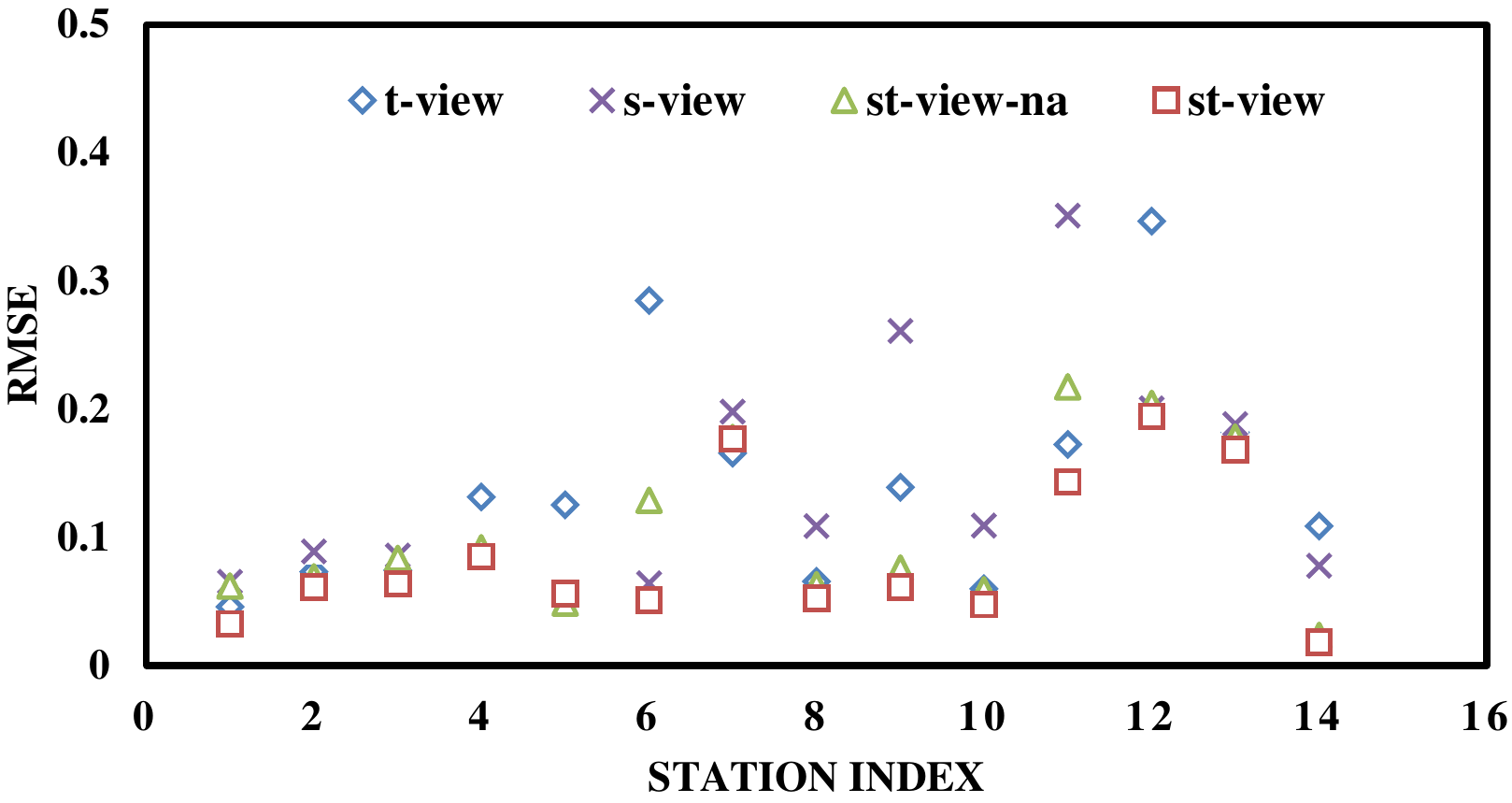}}
  \caption{Comparative predictive performance illustration of stMTMV over other predictive models, its model variants, and the view combinations among stations, respectively.}\label{PerNodeEvaluation}
\end{figure*}

To investigate the strength of stMTMV model among stations, we compared our stMTMV model against other baselines and its variants among different stations. Figure \ref{PerNodeEvaluation} illustrates the predictive performance comparison for stations. It can be seen that stMTMV performs no worse than other baselines as well as its variants for most of the stations. Figure \ref{ModelComparisonNode} presents the comparison of stMTMV over other baselines, where stMTMV is significantly better than other models in at least 6 of the stations. In addition, the results in Figure \ref{ModelComponentNode} demonstrate that stMTMV stably outperforms stMTMV-\emph{us} in all stations, which justifies the importance of spatial correlations among stations and further verifies that the tasks are mutually correlated and the spatial autocorrelation plays an important role in the co-prediction tasks. Moreover, as demonstrated in Figure \ref{ViewComparisonNode}, stMTMV consistently outperforms other view combinations, which, in turn, further provides reasonable justification for the heterogeneous information aggregation.

\subsection{Predictive Accuracy}
Furthermore, we also investigated the predictive accuracy of our stMTMV model with single-task learning and single-view learning methods. The results are demonstrated in Table \ref{PredictiveAccuracyComparison}. The predictive accuracy $Acc$ is computed as follows:
\begin{equation}\label{AccuracyEquation}
	Acc = 1 - \frac{1}{M}\sum_{l=1}^{M} \frac{ \|\textbf{y}_{l} - \hat{\textbf{y}_{l}} \|_{1}}{\| \textbf{y}_{l} \|_{1}},
\end{equation}
where $\textbf{y}_{l}$ is the ground truth and $\hat{\textbf{y}_{l}}$ is the predicted value for the station $l$, and $\|\cdot\|_{1}$ denotes the $\ell_{1}$-norm. From this table, we can have the observations as follows: 1) The predictive accuracy of all methods exhibits a decrease trend for the next 1-4 hours. However, as compared to other methods, the decrease in our method is quite small in magnitude, which demonstrates the robustness of our approach. This further verifies the assumption that the co-prediction among all the stations can boost the performance. 2) Our approach achieves higher accuracy as compared to either single-task or single-view methods, which justifies the effectiveness of combining the multi-task with multi-view algorithms together. 

\begin{table}[!h]
	\begin{center}
		\caption{Predictive accuracy comparison of our model with single-task and single-view methods over next 1-4 hours.}\label{PredictiveAccuracyComparison}
		\renewcommand{\arraystretch}{1.45}
		\begin{tabular}{|c||c|c|c|c|}
			\hline
			\small{Predictive Accuracy}      & 1 hour  &  2 hour    & 3 hour  & 4 hour   \\
			\hline
			\hline
			LR                           & $80.1\%$ &  $76.6\%$  &  $71.4\%$ &  $68.2\%$ \\
			\hline
			\emph{t-view}         & $83.2\%$ &  $77.3\%$  &  $70.6\%$ &  $63.2\%$ \\ 
			\hline
			\emph{s-view}        & $76.0\%$ &  $73.4\%$  &  $71.6\%$ &  $70.8\%$ \\ 
			\hline
			stMTMV                  & $\textbf{85.9\%}$ &  $\textbf{85.0\%}$   & $\textbf{84.3\%}$ &  $\textbf{84.1\%}$ \\
			\hline
		\end{tabular}
	\end{center}
\end{table}

\subsection{Water Quality Predictions}
\begin{figure}[!h]
  \centering
  \includegraphics[width=0.49\textwidth]{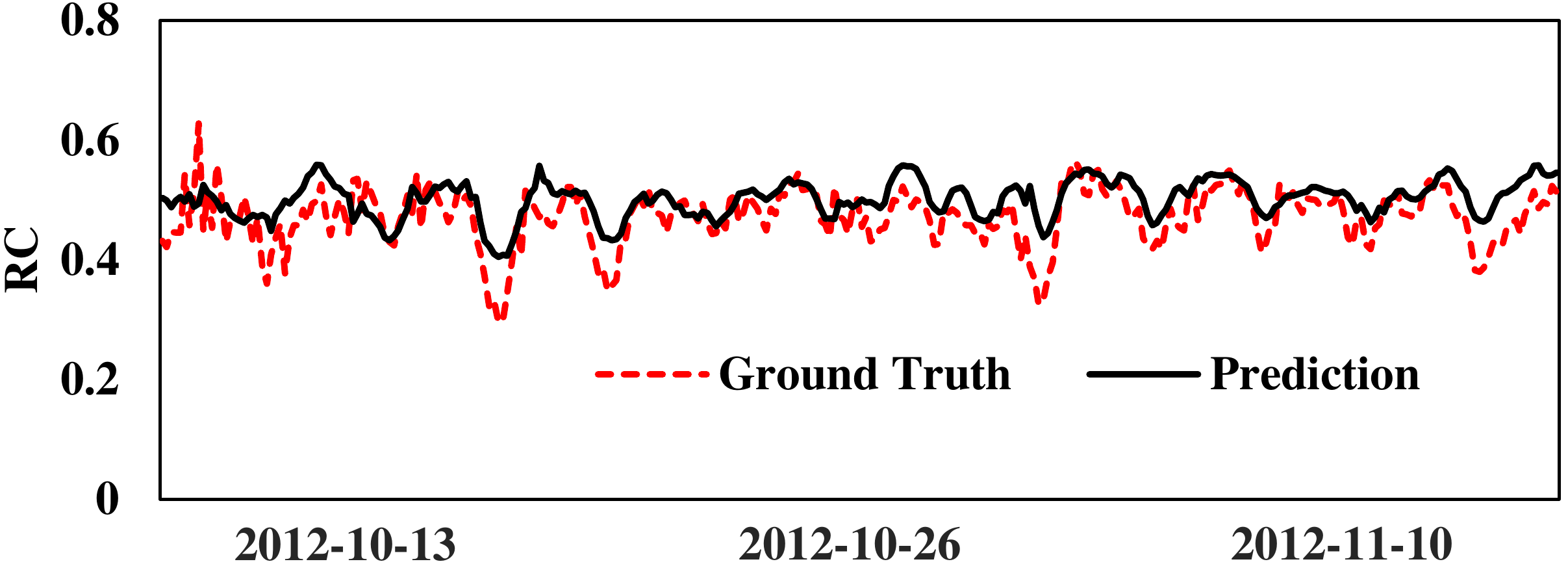}\\
  \caption{Predictions of stMTMV against the ground truths.}\label{PredictionResults}
\end{figure}

Figure \ref{PredictionResults} depicts the predictive results of our method over the next 1 hour against the ground truth in Shenzhen from October 2012 to November 2012. In general, our model is very accurate in tracing the ground truth curves (including sudden changes) of the water quality in Shenzhen City, which demonstrates the effectiveness of our approach.

\section{Related Work}\label{RelatedWork}
\subsection{Classical Model-based Approaches}
It is worth mentioning that several research efforts have been dedicated to model-based approaches for urban water quality prediction~\cite{monteiro2014modeling}\cite{rossman1996numerical}\cite{rossman_RCmodeling_1994}. The main idea behind this kind of approaches is to utilize the first-order or higher-order kinetics to model the chlorine decay along the water distribution system. However, as the mechanisms of the chlorine decay is quite complicated and comprise of reactions with bulk fluid, pipe and natural evaporation, the accurate mathematical modelling of chlorine decay along the water supply system is a tough problem whose solution is not yet absolutely mastered~\cite{castro2003chlorine}. Moreover, the developed decay model requires extensive human labors to perform model calibration with pipe networks, and it depends heavily on the pipe internal surface materials, temperatures, network structure, which makes it difficult to extend to other cities' water distribution systems. Compared to model-based approaches, data-driven based approaches demonstrate their advantages in both flexible and extendibility in many other urban applications~\cite{lee2013understanding}\cite{Zheng_UAirPrediction_KDD2015}\cite{cui2013tracking}\cite{liu2016recognizing}\cite{liu2012fusion}\cite{Zheng_UAir_KDD2013}\cite{Zheng_crossDomain_GIS2015}\cite{Zheng_UrbanComputing_TIST2014}\cite{Zheng_CityNoise_UbiComp2014}, such as urban sensing~\cite{lee2010survey}\cite{lu2016towards}\cite{perttunen2015urban}\cite{zhang2015sensing}, urban air forecast~\cite{Zheng_UAir_KDD2013}\cite{Zheng_UAirPrediction_KDD2015}, destination prediction~\cite{xue2013destination_ICDE2013}\cite{liu2010visual}\cite{Zheng_DataFusion_TBD2015}\cite{zheng2011urban}, and traffic prediction~\cite{wang2014travel_KDD2014}. However, to the best of our knowledge, the literature on urban water quality prediction from the data-driven perspective is relatively sparse.

\subsection{Classical Data-driven based Approaches}
Several studies in the environmental science have been tried to analyze the water quality problems via data-driven based approaches, and those studies covers a range of topics, from the physical process analysis in the river basin, to the analysis of concurrent input and output time series~\cite{solomatine2008data_dataDriven}\cite{haykin2004comprehensive_DD}. The approaches adopted in these studies include  instance-based learning models (e.g., kNN) as well as neural network models (e.g., ANN). In general, those data-driven approaches in the environmental science can fall into the following three major categories: Instance-based Learning models (IBL), Artificial Neural Network models (ANN) and Support Vector Machine models (SVM).

Instance-based learning models (IBL) is a family of learning algorithms that model a decision problem with instances or examples of training data that are deemed important to the model~\cite{aha1991instance_IBL}. As a typical example of IBL, k-Nearest Neighbors (\textit{k}-NN) is widely used due to its simplicity and incredibly good performance in practice. For example, the work introduced by Karlsson~\emph{et al.}~\cite{karlsson1987nearest_DD} addressed the classical rainfall-runoff forecasting problem by k-NN algorithm, and demonstrated promising results. Toth~\emph{et al.}~\cite{toth2000comparison_DD} used k-NN to predict the rainfall depths from the history data, and showed the persistent outperformance of k-NN over other time series prediction methods. As another example, Ostfeld~\emph{et al.}~\cite{ostfeld2005hybrid_DD} developed a hybrid genetic k-Nearest Neighbor algorithm to calibrate the two-dimensional surface quantity and water quality model.

Artificial Neural Network (ANN) is a network inspired by biological neural networks (in particular the human brain), which consists of multiple layers of nodes (neurons) in a directed graph with each layer fully connected to the next one~\cite{haykin2004comprehensive_DD}. Neural networks have been widely employed to solve a wide variety of tasks, and can acheive good results. For instance, Moradkhani~\emph{et al.}~\cite{moradkhani2004improved_DD} proposed an hourly streamflow forecasting method based on a radial-basis function (RBF) network and demonstrated its advantages over other numerical prediction methods. Also, the work introduced by Kalin~\cite{kalin2010prediction} predicted the water quality indexes in watersheds through ANN.

Support Vector Machines (SVMs) are typical supervised learning models that analyze data used for classification and regression~\cite{suykens1999least_DD}. In aquatic studies, it was also extended to solving prediction problems~\cite{solomatine2008data_dataDriven}. For instance, Liong~\emph{et al.}~\cite{yu2006support_DD} addressed the issue of flood forecasting using Support Vector Regression (SVR) which is an extension of SVM. Another work by Xiang \emph{et al.}~\cite{yunrong2009water} utilized a LS-SVM model to deal with the water quality prediction problem in Liuxi River in Guangzhou. 

However, none of these approaches is applied into urban scenarios, which is quite different from our applications. Moreover, those existing approaches process the data from a single source, and can hardly integrate the data from different sources. Thus, their applications in the urban scenarios are restricted.

\subsection{Multi-task Multi-view Learning Approaches}
Multi-task learning is a learning paradigm that jointly learns multiple related tasks and has demonstrated its advantages in many urban applications, such as transportation and event forecasting~\cite{YeJieping_KDD2015_MTLSpatial}\cite{zheng2013time_MTL_AAAI2013}\cite{Zheng_UrbanComputing_TIST2014}. In particular, it is more effective in handling those with insufficient training samples~\cite{Liu_IJCAI2015_MultiTask}\cite{ZhangYu_UAI_MTL_2010}\cite{zhang_IJCAI2013_MTLHighOrderLearning}\cite{chen2011integrating_MTL_KDD2011}\cite{liu2009multi}\cite{gong2012robust}. However, most of the existing approaches only explore the task relatedness, but ignore the consistency information among different views within a task. 

Multi-view learning has been proposed to leverage the information from diverse domains or from various feature extractors, and combining the heterogeneous properties from different views can better characterize objects and achieve promising performance~\cite{bulling2012multimodal}\cite{Muslea_MultiView_IJCAI2003}\cite{zhang2013multi_IJCAI2013}\cite{Zheng_DataFusion_TBD2015}. Nevertheless, existing multi-view learning approaches discard the label information from other related tasks, which usually leads to suboptimal performance. 

Thus, multi-view multi-task learning is proposed to explore both task relatedness and view relatedness simultaneously within a learning framework~\cite{he2011graph_IteM2_ICML2011}\cite{jin2013shared_CSL_MTVT}\cite{liu2016action_MTL}\cite{zhang2012inductive_regMVMT_KDD2012}\cite{MVMTL_AAAI2016_Career}\cite{MVMTL_IJCAI2016_UrbanWater}. For example, He \emph{et al.}~\cite{he2011graph_IteM2_ICML2011} proposed a graph-based iterative framework ($GraM^{2}$) for multi-view multi-task learning and obtained impressive results in text categorization applications.
However, as far as we know, the literature on spatio-temporal based multi-task multi-view learning is relatively sparse. To the best of our knowledge, our approach is the first work on spatio-temporal based multi-task multi-view learning, which can incorporate spatio-temporal heterogeneities via a multi-task multi-view learning framework and is able to applied to other spatio-temporal based applications.

\section{Conclusion and Future Work}\label{Conclusion}
This paper presents a novel data-driven approach to forecast the water quality of a station by fusing multiple sources of urban data.  We evaluate our approach based on Shenzhen's water quality and various urban data. The experimental results demonstrate the effectiveness and efficiency of our approach. Specifically, our approach outperforms the traditional RC decay model~\cite{rossman_RCmodeling_1994} and other classical time series predictive models (ARMA, Kalman) in terms of RMSE metric. Meanwhile, as our approach consists of two components, each of the components demonstrates its effectiveness through extensive experiments and analysis. In particular, the first component is the influential factors identification, which explores the factors that affect the urban water quality via extensive experiments and analysis in Section~\ref{TemporalFeature} and~\ref{SpatialFeature}. The second one is a spatio-temporal multi-view multi-task learning (stMTMV) framework that consists of multi-view learning and multi-task learning. The experiments have shown that stMTMV has a predictive accuracy of around $85\%$ for forecasting next 1-4 hours, which outperforms the single-task methods (LR) by approximately $11\%$ and the single-view methods (\emph{t-view} and \emph{s-view}) by approximately $11\%$ and $12\%$, respectively. The code has been released at: https://www.microsoft.com/en-us/research/publication/urban-water-quality-prediction-based-multi-task-multi-view-learning-2/

In future, we plan to deal with the water quality inference problems in the urban water distribution systems through a limited number of water quality monitor stations.

\section*{Acknowledgments}\label{Acknowledgement}
This work was supported by the China National Basic Research Program (973 Program, No. 2015CB352400), NSFC under grant U1401258, NSCF under grant No. 61572488. This research was also supported in part by grants R-252-000-473-133 and R-252-000-473-750 from the National University of Singapore. We also thank Yipeng Wu for sourcing the data in this study.

\bibliographystyle{IEEEtran}
\bibliography{references}

\ifCLASSOPTIONcaptionsoff
  \newpage
\fi

\end{document}